\documentstyle[twocolumn,prc,aps,epsf,epsfig]{revtex}

\newcommand{\po}{\partial_\omega}
\newcommand{\pt}{\partial_\tau}
\newcommand{\dn}{{\rm dn}}

\newcommand{\IA}{$\bar I I \,$}
\begin{document}
\twocolumn[\hsize\textwidth\columnwidth\hsize\csname@twocolumnfalse\endcsname
\preprint{SUNY-NTG-xxx}
\preprint{NT@UW-02-008}

\title{
Forced Tunneling and Turning State Explosion in Pure Yang-Mills Theory 
}

\author{D. M. Ostrovsky$^1$, G. W. Carter$^2$, and E. V. Shuryak$^1$ }

\address{
$^1$Department of Physics and Astronomy, State University of 
New York, Stony Brook, NY 11794-3800\\
$^2$Department of Physics, Box 351560, University of Washington, Seattle,
WA 98195-1560
}

\date{18 April 2002}
\maketitle

\begin{abstract}
We consider {\em forced} tunneling in QCD, described semiclassically by
instanton-antiinstanton field configurations.
By separating topologically different minima we obtain details of the 
effective potential and study the {\em turning states}, which are similar
to the sphaleron solution in electroweak theory.
These states are alternatively derived as minima of the
energy under the constraints of fixed size and Chern-Simons number.
We study, both analytically and numerically, the subsequent evolution of 
such states by solving the classical Yang-Mills equations in real time, 
and find that the gauge field strength is quickly localized into an
expanding shell of radiating gluons.
The relevance to high-energy collisions of hadrons and
nuclei is briefly discussed.
\end{abstract}
\vskip 2pc]

\section{Introduction}

\subsection{Instanton-Induced Scattering in QCD}

The existence of topologically distinct non-abelian
gauge fields, with tunneling
between corresponding classical vacua described semiclassically by
instantons \cite{BPST}, is one of the most spectacular nonperturbative
effects of field theory.
Significant progress has been made in understanding instanton-induced
effects in Quantum Chromodynamics (QCD), 
explaining both explicit U$_A$(1) chiral symmetry
breaking at the single-instanton level \cite{tHooft} and spontaneous
SU($N_f$) chiral symmetry breaking by the instanton 
ensemble \cite{inst_chiral}. 
Euclidean correlation functions, studied phenomenologically and on the 
lattice, have been explained to a significant extent by instantons as well
\cite{SS_98}.

With tunneling phenomena apparently so important in {\em virtual}
quark and gluon propagation, it is reasonable to think them
also relevant in {\em real} processes such as scattering or 
particle production in Minkowski space.
We thus seek contributions to parton scattering amplitudes from 
the theory of instanton-related objects, and supporting experimental evidence.

With this as our motivation, we concentrate in this paper on the
theoretical basis of such effects from pure Yang-Mills theory.
Specific applications to high-energy processes with hadrons or nuclei
are left for papers to follow, although we will discuss phenomenological
generalities where relevant.

Progress in understanding of the role of tunneling in
high energy processes has been tempered by technical problems for years.
Significant insights were obtained in the 1980's \cite{electroweak80s}
and further developed in the early 1990's \cite{electroweak90s,KhR} 
through work in electroweak theory.
In this case, the instanton-induced cross section is readily identified by 
baryon number violation and many noteworthy features of these processes 
were found.
However, quantitative estimates of the associated cross sections
proved to be far below observable limits and interest quickly waned.
Similar ideas have also been developed in QCD \cite{MRS}, 
notably the search for hard 
processes induced by {\em small-sized instantons} which continues
at HERA \cite{RS}.

Another role for instanton-induced processes has recently been proposed by
Kharzeev, Kovchegov, and Levin \cite{KKL} and
Nowak, Shuryak, and Zahed \cite{NSZ}.
These works focus on typical QCD instantons, of size $\rho\sim 1/3$ fm
\cite{inst_chiral}, which determine the {\em semi-hard} scale 
of $Q\sim 1-2$ GeV. 
It was proposed  that topological tunneling is behind the well-known 
features of high energy scattering described phenomenologically by the 
so-called ``soft'' pomeron.
These ideas were further tested in Ref.~\cite{COS}, where they were
demonstrated to be reasonably consistent with experimental data.

\begin{figure}[ht]
\begin{center}
  \begin{minipage}[c]{1.8in}
    \centering
     \includegraphics[totalheight=1.8in]{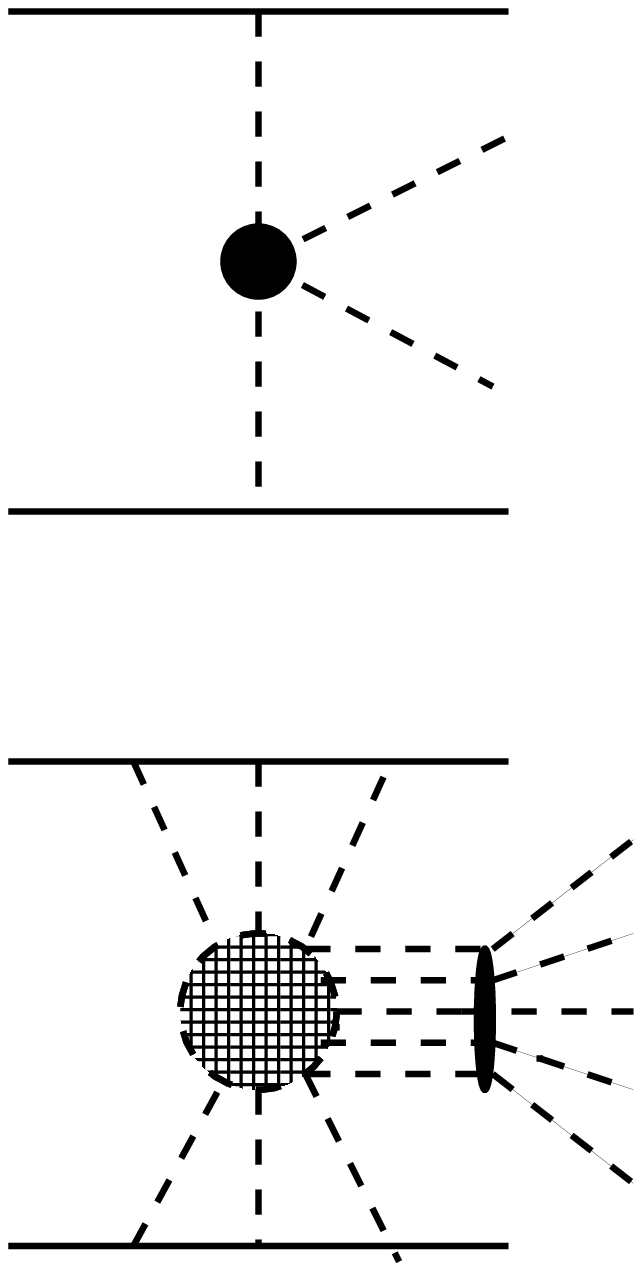}
   \end{minipage}
\vskip 5mm 
  \begin{minipage}[c]{2.6in}
    \centering
      \includegraphics[width=2.6in]{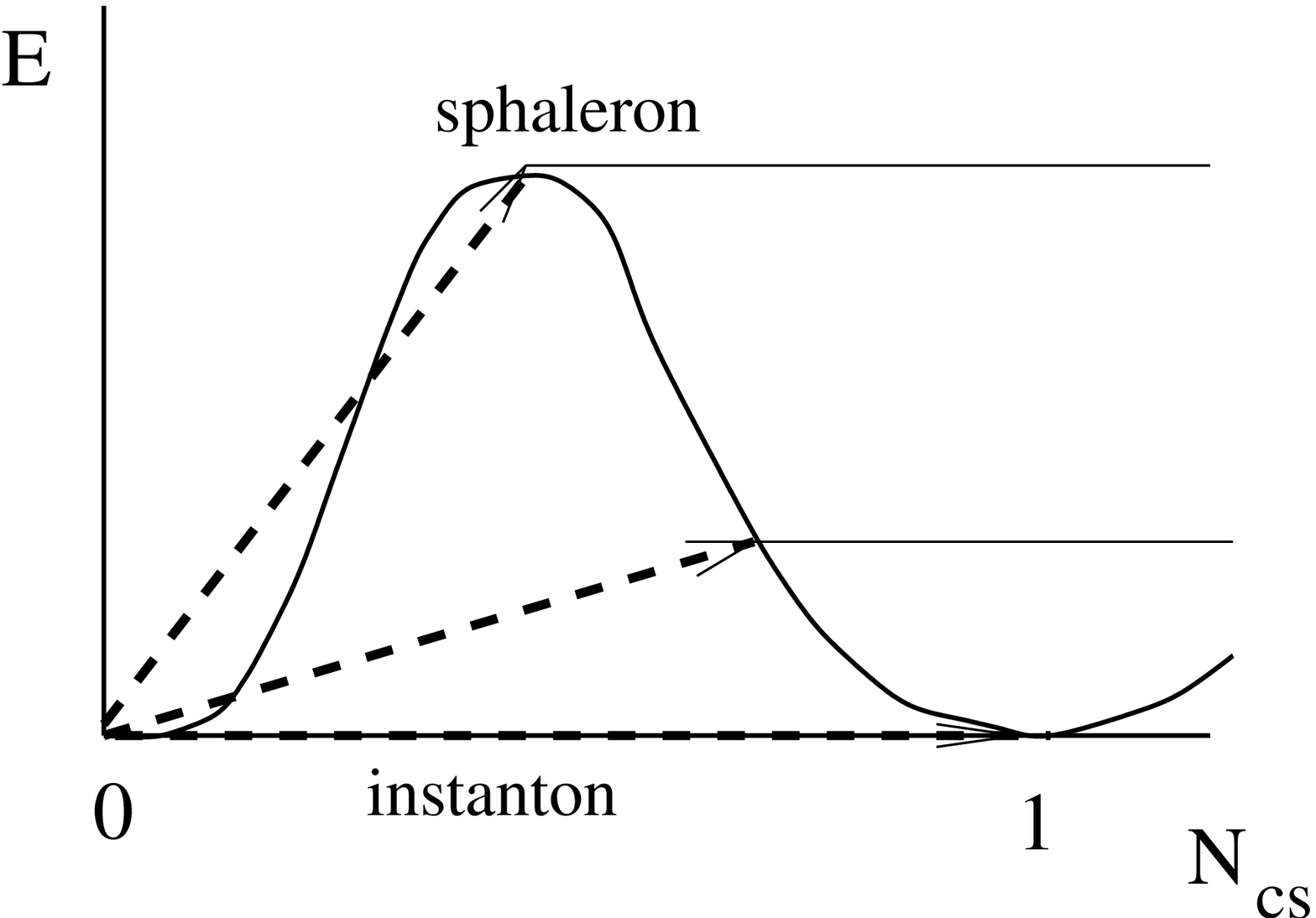}
   \end{minipage}
\end{center}
  \caption[]
  {
   \label{processes}
The top of the figure compares
(a) a typical inelastic perturbative process (two t-channel gluons collide,
producing
a pair of gluons) to (b) a nonperturbative inelastic process, incorporating
collisions
of a few $t$-channel gluons with the instanton (the shaded circle),
resulting in multi-gluon production.
The bottom figure (c) shows the same process, but in a 
quantum mechanical way.  The energy of Yang-Mills field versus
the Chern-Simons number, $N_{cs}$, is a periodic
function, with zeros at integer points. The {\em instanton} (shown by the 
lowest dashed line) is a transition between such points. However if some nonzero
energy is deposited into the process during transition, the virtual path
(the dashed line) leads to a  {\em turning states}, from which starts the
real time motion outside the barrier (shown by horizontal solid lines).
The maximal cross section corresponds to the transition to
the top of the barrier, called a {\em sphaleron}.
 }
\end{figure}

Since the 1960's attempts have been made to explain high-energy hadronic
collisions with multi-peripheral models, with various ladder diagrams
describing hadron production. 
It was realized that in order to get cross-sections which are not falling at 
high energies, one needed $vector$ field exchange in the $t$-channel. 
With the discovery of QCD, gluons naturally play this role. 
Generic pQCD-inspired models appeared with processes like that
shown in Fig.~\ref{processes}(a).
Eventually this development led to the BFKL gluon ladder \cite{BFKL}, 
which produces an (approximately) supercritical pomeron, a
``hard'' pomeron with the intercept well above 1. 
Recent studies of high energy hard processes, especially at HERA, 
have indeed found strong growth of the cross section with energy for truly 
hard processes ($Q^2 \gg 1$ GeV$^2$), consistent with the BFKL
treatment.

But various data at the {\em semi-hard} scale of $Q^2 \sim 1$ GeV$^2$
demonstrate rather different growth with energy, consistent with a
``soft'' pomeron.
Whatever it might be, the pomeron should be an object of a particular size
deduced from the slope of its Regge trajectory,
$\alpha'\sim 1/(2 \, {\rm GeV})^2$. 
This size of course cannot be explained by basically scale-invariant pQCD,
and thus calls for a nonperturbative derivation.

Existing models for the soft pomeron also include ladders made of $t$-channel
gluons, and the differences between them lie mainly in the construction of
their rungs. 
Each of the various models has a unique answer for
{\em what is actually produced} in gluon-gluon partonic collisions.
For example, in Ref.~\cite{KL} a pair of pions in the scalar channel 
or a scalar glueball is produced.
The introduction into this problem of instanton-induced vertices
\cite{Shu_toward,KKL,NSZ}, shown schematically in Fig.~\ref{processes}(b),
led to a different idea: the object produced is neither a gluon (as in BFKL)  
nor any {\em colorless} hadronic state, but rather a {\em colored} cluster 
of the gluon field, which in turn decays into several gluons. 
It has been shown that the cross section peaks at an invariant cluster mass
in the range $2.5-3$ GeV \cite{KKL,NSZ}.
It is very important that the states which are produced are not a random
group of gluons, but rather their coherent superposition.
Understanding their composition is the main objective of this work.

A quantum-mechanical interpretation of the collision process is central 
to this question of prompt gluon production.  
An impressive body of work has addressed this problem with classical 
Weizs\"acker-Williams fields of gluons, the Color Glass Condensate
\cite{cgc}.
Here we consider a different classical process, one involving topological
objects.
In Fig.~\ref{processes}(c) we schematically show a barrier separating 
two topologically distinct classical vacua, with
Chern-Simons numbers\footnote{This will be introduced formally below. Here it
is sufficient to note only that we consider a definite pair of gauge
potentials, separated on one of the many coordinates of our quantum system.} 
$N_{CS}=$ 0 and 1. 
Unlike a standard instanton transition, shown by the horizontal dashed line, 
in a high energy collision a finite amount of energy is absorbed.
This can be viewed as a ``forced tunneling'' event (either of the other 
two dashed lines) which ends at a {\em turning state}, where the total 
energy is equal to the potential energy, so that the paths can exit the 
(Euclidean) domain below the barrier.
These colored unstable objects are close relatives of electroweak 
sphalerons \cite{Manton,KM}, which are defined at the barrier's peak. 
We will demonstrate how these objects then evolve with conserved energy, 
developing into an exploding shell of color field.
This part of the process is diagrammed with the horizontal lines in
Fig.~\ref{processes}(c).

Before we come to these explosions, we will discuss in detail the
instanton--anti-instanton (\IA) configurations which describe this
forced tunneling. 
They provide one way toward the understanding of the 
effective potential separating topologically different gauge fields, 
as well as the turning states themselves.
We then proceed to another derivation of the same results as static 
solutions in classical Yang-Mills theory constrained in size.
The real-time decay of the static configuration is studied in detail, using 
both analytic and numerical methods, ultimately leading to a description
of the expanding shells in terms of gluonic quanta.

\subsection{Spherically Symmetric Yang-Mills Fields}

For the SU(2) color subgroup in which we are interested, spherically 
symmetric field
configurations of the gauge field ${\cal A}_\mu^a$ can be expressed 
through the following four space-time (0, $j=1..3$) 
and color ($a=1..3$) structures 
\begin{eqnarray}
\label{sph_ansatz}
{\cal A}^a_j &=& A(r,t)\Theta^a_j + B(r,t) \Pi^a_j + C(r,t)\Sigma^a_j
\nonumber\\
{\cal A}^a_0 &=& D(r,t) \frac{x^a}{r}
\end{eqnarray}
with
\begin{equation}
\Theta^a_j=\frac{\epsilon_{jam}x^m}{r}\,, \quad \Pi^a_j =
\delta_{aj}-\frac{x_ax_j}{r^2}\,, \quad \Sigma^a_j = \frac{x_ax_j}{r^2}\,.
\label{projectionops}
\end{equation}
It is convenient to express the scalar functions in Eq.~(\ref{sph_ansatz}) 
in terms of four $r$ and $t$ dependent functions, which are similar to the 
fields of the 1+1 dimensional Abelian gauge-Higgs model
($A_{\mu=0,1},\,\phi,\, \alpha$)
on a hyperboloid \cite{Witten_ans}:
\begin{equation}
A=\frac{1+\phi\sin\alpha}{r}, \quad B=\frac{\phi\cos\alpha}{r},\quad
C=A_1,\quad D=A_0.
\end{equation}

One can express the field strengths in these terms as
\begin{eqnarray}
{\cal E}^a_j = {\cal G}^a_{0j} 
&=&\frac{1}{r} [\partial_0\phi\sin\alpha +
\phi\cos\alpha(\partial_0\alpha-A_0)]\Theta^a_j \nonumber\\
&&+ \frac{1}{r} [\partial_0\phi\cos\alpha -
\phi\sin\alpha(\partial_0\alpha-A_0)]\Pi^a_j
\nonumber\\&&
+ (\partial_0A_1-\partial_1A_0)\Sigma^a_j
\label{electric}\end{eqnarray}
and
\begin{eqnarray}
{\cal B}^a_j = \frac{1}{2}\epsilon_{jkl}{\cal G}^a_{kl} 
&=&\frac{1}{r} [-\partial_1\phi\cos\alpha +
\phi\sin\alpha(\partial_1\alpha-A_1)]\Theta^a_j \nonumber\\
&&+ \frac{1}{r} [\partial_1\phi\sin\alpha +
\phi\cos\alpha(\partial_1\alpha-A_1)]\Pi^a_j 
\nonumber\\ &&
+ \frac{1-\phi^2}{r^2}\Sigma^a_j, 
\label{magnetic}\end{eqnarray}
where $\partial_0\equiv \partial_t \mbox{ and } \partial_1\equiv\partial_r$.

The action in 3+1 dimensional Minkowski $(-,+,+,+)$ space reduces as
\begin{eqnarray}\label{action}
S &=& \frac{1}{4g^2}\int d^3x dt\left[\left({\cal B}^a_j\right)^2 
- \left({\cal E}^a_j\right)^2\right]
\nonumber\\
&=& 4\pi\int dr dt \Bigg[\left(\partial_\mu\phi\right)^2+
\phi^2\left(\partial_\mu\alpha-A_\mu\right)^2\nonumber\\
&&+\frac{(1-\phi^2)^2}{2r^2}-
\frac{r^2}{2}\left(\partial_0A_1-\partial_1A_0\right)^2\Bigg] \,,
\end{eqnarray}
with the summation now over the 1+1 dimensional $(-,+)$ metric.

The spherical ansatz is preserved by a set of gauge transformations
generated by unitary matrices of the type
\begin{equation}
\label{gauge_transform}
U(r,t)=\exp\left(i\frac{\beta(r,t)}{2r}\tau^ax^a\right).
\end{equation}
These transformations naturally coincide with the gauge symmetry of 
the corresponding abelian Higgs model:
\begin{equation}
\phi'=\phi, \quad \alpha'=\alpha+\beta, \quad A'_\mu=A_\mu+\partial_\mu
\beta\,.
\end{equation}
This freedom can be used to gauge out, for example, the 
${\cal A}_0$ component.

Topological properties of the gauge field are governed by the
topological current
\begin{equation}
K_\mu=-{1 \over 32 \pi^2} \epsilon^{\mu\nu\rho\sigma}
\left({\cal G}^a_{\nu\rho} {\cal A}^a_\sigma -
{g \over 3}\epsilon^{abc} {\cal A}^a_\nu  {\cal A}^b_\rho  {\cal A}^c_\sigma
\right)\,.
\end{equation}
Although this current is not gauge invariant, its change is
related to the (gauge invariant) local topological charge
\begin{equation}
\partial_\mu K^\mu
= -{1 \over 32 \pi^2} {\cal G}^a_{\mu\nu}\tilde {\cal G}^a_{\mu\nu} \,.
\end{equation}
Within the spherical ansatz and the ${\cal A}_0=0$ gauge the topological
current takes a simpler form,
\begin{eqnarray}
K^0 &=& \frac{1}{8\pi^2r^2}
\left[(1-\phi^2)(\partial_1\alpha-A_1)-\partial_1(\alpha-\phi\cos\alpha)\right]
\nonumber\\
K^i &=& \frac{x^i}{8\pi^2r^3}
\left[(1-\phi^2)\partial_0\alpha-\partial_0(\alpha-\phi\cos\alpha)\right] \,,
\end{eqnarray}
while the topological charge becomes
\begin{eqnarray}
\partial_\mu
K^\mu &=& \frac{1}{8\pi^2r^2}\left\{-\partial_0\left[(1-\phi^2)
(\partial_1\alpha-A_1)\right]\right.
\nonumber\\
&&+\left.\partial_1\left[(1-\phi^2)(\partial_0\alpha-A_0)\right]\right\}\,.
\end{eqnarray}
Note that only gauge-invariant combinations of field derivatives 
appear here.

As a ``topological coordinate'' marking the tunneling paths and
the turning states one can use the Chern-Simons number
\begin{eqnarray}
N_{CS} = \int d^3x K_0
&=& -\frac{1}{2\pi}\int dr
(1-\phi^2)(\partial_1\alpha-A_1)
\nonumber\\&&
+\frac{1}{2\pi} \left.(\alpha-\cos\alpha)\right|_{r=0}^{r=\infty}
\label{csdef}
\end{eqnarray}
The first, gauge-invariant term is sometimes called
the {\em corrected} or {\em true} Chern-Simons number \cite{AKY,FKS},
$\tilde{N}_{CS}$, while the second (gauge-dependent) term is referred to as
the {\em winding number}. It is the change in $\tilde{N}_{CS}$ which
is equivalent to the integral over the local topological charge.

\section{ Instanton-Antiinstanton Configurations}

\subsection{Forced Tunneling}

A brief introduction to the quantum mechanics of gluons in high energy
collisions has been given in the introduction.
The effect of colliding partons can be included in various forms.
For example, these fields can be represented as {\em non-zero}
external currents which affect the tunneling paths of Yang-Mills field. 
In the zero-current, vacuum case, the usual instanton solutions
are spherically symmetric in four Euclidean dimensions. 
The collision problem of two (or more) partons, on the contrary, at
non-zero impact parameters does not have even an {\em axial} symmetry.
The reader therefore may wonder why this (and all previous works)
on the subject consider 3+1 dimensional spherically symmetric fields.

The justification for this ansatz is that the absolute magnitude of the 
tunneling field is large compared to external forces.
Also, as will be shown below, 
spherically symmetric clusters are an energy minimum
for fixed size and topological coordinate. 
Should the resulting cluster not have exact spherical symmetry
one can always approach the problem perturbatively, considering first the 
external forces projected onto the direction of tunneling, and then other
components as small corrections.
The resulting 1+1 dimensional problem is readily solved numerically and,
to a great extent, analytically.

Unlike separated instantons ($I$) and antiinstantons ($\bar I$), 
combined \IA configurations are neither selfdual nor anti-selfdual
and do not satisfy classical equations of motion.  
They are not extrema of the action, since they describe the valley 
stretching between true extrema -- the zero field (equivalent to
an \IA at zero separation) and well-separated \IA pair.
Substituting any \IA trial function into the of Yang-Mills equation
of motion, we find a finite
\begin{equation} 
D_\mu {\cal G}_{\mu\nu}=J_\nu  \,.
\end{equation} 
This means some external current must be applied to the gauge fields if we 
want to use semiclassical analysis.
The process can only then be interpreted as a classical \IA, or a
{\em forced path}.
There are two interpretations of \IA configurations with different
consequences.

The historical view is that such fields describe quantum fluctuations in 
the Yang-Mills vacuum, the process in which a virtual path goes
under the barrier, then reverses course and ends up in the {\em same} 
minimum from which it started.
This process has zero net topological charge.
Naturally, the early studies concentrated on the action corresponding 
to these configurations, the quantity which controls its weight in the 
path integral.
The first such work was done long ago by Callan, Dashen, and Gross
\cite{CDG}, resulting in a dipole force and the
action $\delta S \sim 1/T^4$ at large distance $T$ between the centers.
Higher terms in the multipole expansion have been discussed
in literature after that, e.g. \cite{yung}. 
When it was eventually realized that quark-induced \IA pairings are more 
important for the instanton ensemble in QCD \cite{SS96}, interest in the 
pure Yang-Mills theory waned.

In this paper we will however take a different view of \IA configurations.
Since the external forces from the partonic current do work on the \IA
pair, the energy at intermediate times is non-zero. 
We will consider only cases in which the fields at positive and negative times
are essentially the same (modulo a sign and, sometimes, a
gauge transformation).  
Thus this energy will be even under $t\rightarrow -t$, with a
natural maximum at $t=0$. 
As a result, all quantities which are odd under this transformation
(like the electric field) naturally vanish at this instant.
The remaining, {\em purely magnetic} configuration
is what we define as the {\em turning state} of this path.

The resulting action corresponds to an excitation {\em probability}
of this turning state created by the external current $J$,
\begin{equation} 
P \sim \left| \langle 0|J|{\rm turning \, state}\rangle \right|^2 .
\end{equation} 
Through this mechanism the excitation of \IA pairs leads to the production
of real particles, as advertized in the Introduction and to be analyzed 
in the next sections.

\subsection{Simple \IA Trial Functions}

We now consider the simplest example of a possible turning state, a
straightforward \IA sum ansatz.
With it, we can demonstrate some basic features, although we will find
them insufficient for our purposes and move to a more complicated
ansatz in the next subsection.

Written the singular gauge, the {\em sum ansatz} is:
\begin{equation} 
\frac{g}{2} {\cal A}_{a\mu}^{sum}(x)= 
\frac{ \bar \eta_{a\mu\nu} y_1^\nu \rho^2}{ y_1^2(y_1^2+\rho^2)}+ 
\frac{ \eta_{a\mu\nu} y_2^\nu \rho^2 }{y_2^2(y_2^2+\rho^2)}\,,
\end{equation} 
where we assume that both the instanton and the antiinstanton (the first 
and second terms, respectively) have the same color orientation and size
$\rho$. 
The vectors $y_1=x-z_I$ and $y_2=x-z_{\bar I}$ are the distances from
the observation point $x$ to the instanton and antiinstanton centers. 
In what follows we assume $z_I=(T/2,0,0,0)$ and $z_{\bar I}=(-T/2,0,0,0)$, 
where the imaginary time between centers is $T$.

Note that although a single instanton's profile behaves as $1/x$ near the
origin, the physical quantity $\left({\cal G}^a_{\mu\nu}\right)^2$ is finite.
However, for the sum ansatz this feature is lost and the same quantity 
goes as $1/x^2$ near the origin.

This unphysical feature can be quickly remedied by the {\em ratio 
ansatz} \cite{Shu_rat}, which for identical sizes and orientations is
\begin{equation} 
\frac{g}{2} {\cal A}_{a\mu}^{ratio}(x) = \frac{\eta_{a,\mu\nu}y_1^{\nu}
\frac{\rho^2}{y_1^2} +\bar\eta_{a,\mu\nu}y_2^{\nu}
\frac{\rho^2}{y_2^2} }{ 1+ \frac{\rho^2}{y_1^2}+\frac{\rho^2}{y_2^2} }
\end{equation} 
These trial functions are simple enough to have analytic
expressions for the field strength, the energy of static turning states,
and the Chern-Simons number. 
For reference, one has the following
expressions for the magnetic and electric fields squared:
\begin{eqnarray} 
\vec {\cal B}^2 &=& 16384(768 t^8 + 1024 r^2 t^6 + 3072 t^6 + 2304 t^6 R^2 
\nonumber\\ && 
+ 6400 r^2 t^4 R^2 + 2048 r^2 t^4 + 512 r^4 t^4 + 1824 t^4 R^4
\nonumber \\ && 
+ 3072 t^4 + 4608 t^4 R^2 + 1024 r^6 t^2 + 192 t^2 R^4
\nonumber \\ && 
+ 512 r^2 t^2 R^2 - 1024 r^4 t^2 + 144 t^2 R^6 + 1216 r^2 t^2 R^4 
\nonumber \\ && 
+ 2816 r^4 t^2 R^2 + 288 r^4 R^4 + 768 r^8 + 768 r^6 R^2 
\nonumber \\ && 
+ 48 r^2 R^6 + 3 R^8)/(16 r^4 + 32 r^2 t^2 + 8 r^2 R^2 + 16 t^4 
\nonumber \\ && 
- 8 t^2 R^2 + R^4 + 32 r^2 + 32 t^2 + 8 R^2)^4 \,,
\end{eqnarray}
\begin{eqnarray} 
\vec {\cal E}^2 &=& 1048576 t^{2}(32 r^2 t^4 + 48 t^4 R^2
+ 64 r^4 t^2 + 64 r^2 t^2 \nonumber \\ 
&& + 80 r^2 t^2 R^2
+ 48 t^2 R^2 + 24 t^2 R^4 + 12 R^4 + 32 r^2 \nonumber \\  
&& + 32 r^6 + 64 r^2 R^2 + 26 r^2 R^4 + 64 r^4 R^2 \nonumber \\
&& + 64 r^4 + 12 R^2 + 3 R^6)/
(16 r^4 + 32 r^2 t^2 + 8 r^2 R^2 \nonumber \\ 
&& + 16 t^4 - 8 t^2 R^2 + R^4 + 32 r^2 + 32 t^2 + 8 R^2)^4 \,.
\end{eqnarray} 
Their scalar product is
\begin{eqnarray} 
\vec {\cal B}\cdot\vec {\cal E} &=& -393216 t R(R^2 + 2 + 4 r^2 + 4 t^2)(16 t^4 
\nonumber \\ && 
+ 24 t^2 R^2 + 32 r^2 t^2 + 32 t^2 + R^4 + 16 r^4 
\nonumber \\ && 
+ 8 r^2 R^2)/ (16 r^4 + 32 r^2 t^2 + 8 r^2 R^2 + 16 t^4 
\nonumber \\ && 
- 8 t^2 R^2 + R^4 + 32 r^2 + 32 t^2 + 8 R^2)^4 \,,
\end{eqnarray} 
where we have set $\rho=1$ and $R=T$ is the intercenter distance. 

\begin{figure}[tb]
\hspace*{-5mm}
\begin{center}
\epsfig{file=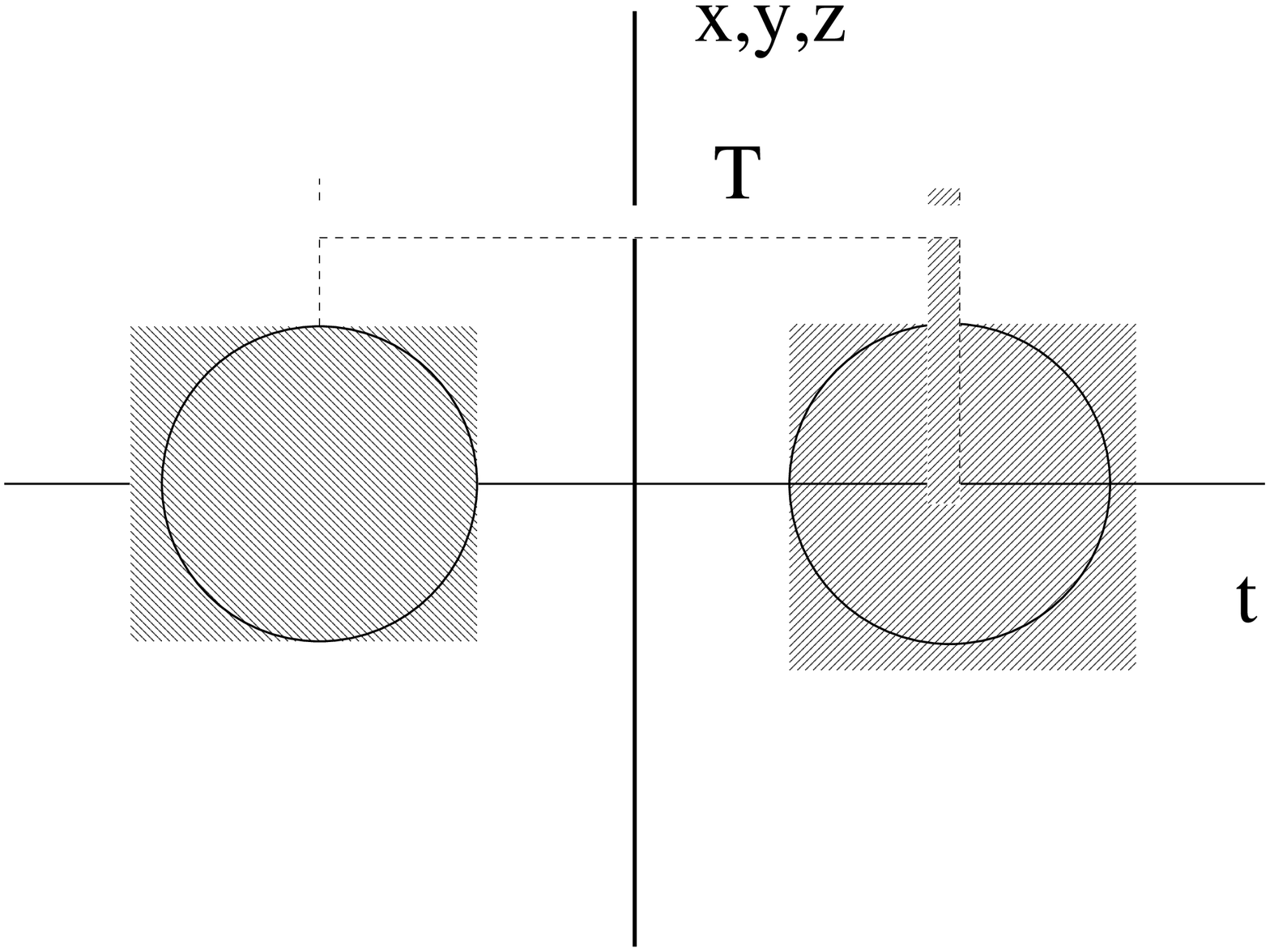, width=45mm}
\vskip 5mm 
\epsfig{file=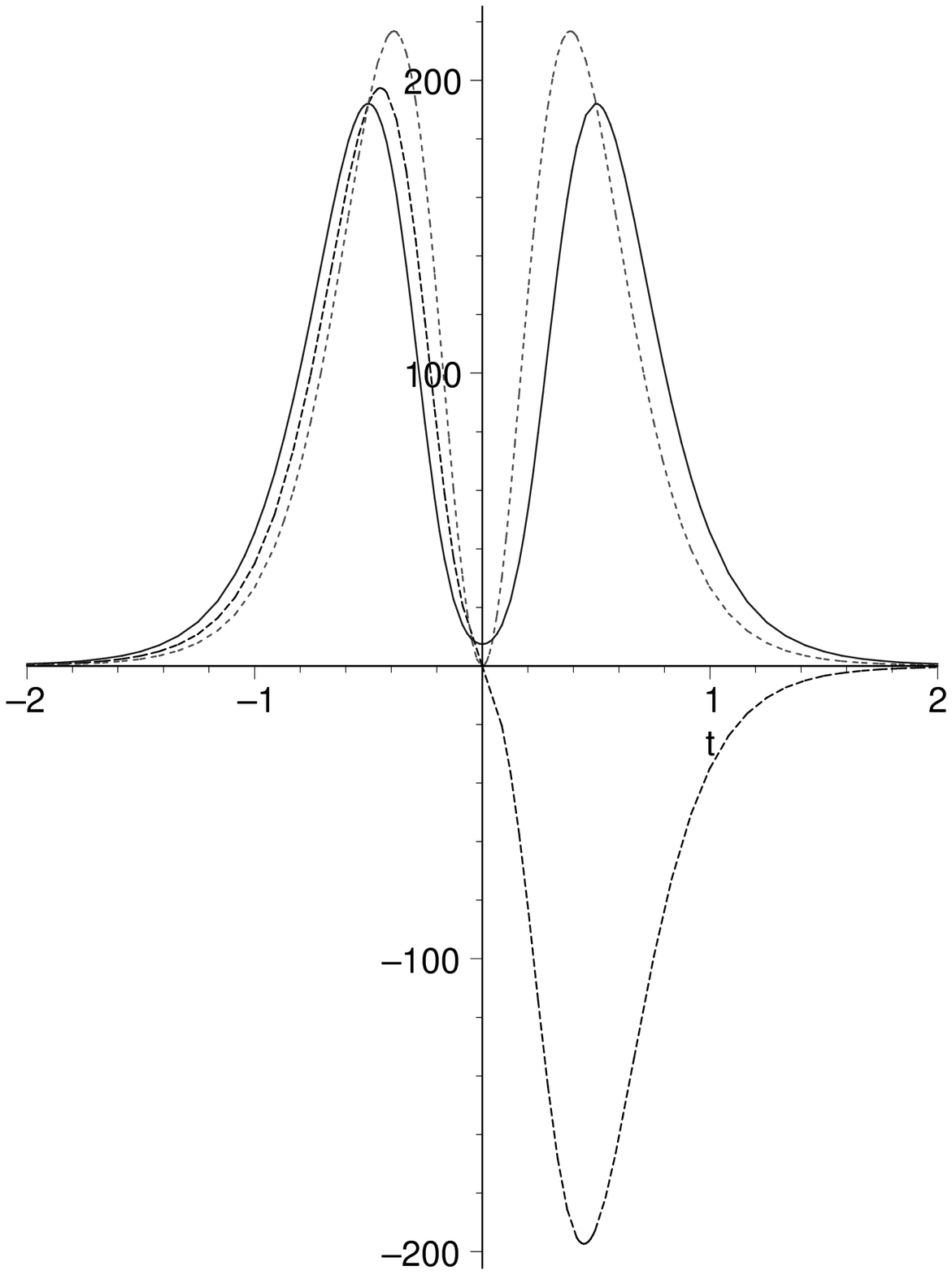, width=50mm}
\end{center}
\caption {\label{IAfig}
Instanton-antiinstanton configurations. (a) A schematic picture in 
Euclidean space-time. The thick vertical line, $t=0$, corresponds to the
location of the turning state. The definition of
the inter-center distance $T$ is also shown. 
(b) Distribution along the time axis
of $2\vec {\cal B}^2$,$2\vec {\cal E}^2$, and 
$2\vec {\cal B}\cdot \vec {\cal E}$ 
for the ratio ansatz with $T=\rho$,
shown by the solid, dashed, and short-dashed lines respectively.
The curve for $\vec {\cal B}\cdot\vec{\cal E}$ is the only one which is $t$-odd.
}
\end{figure}

One can see that, in the simplest case of identical sizes and orientations
for the $I $ and $\bar I$, time reflection symmetry $t\rightarrow -t$ 
of the problem is indeed manifest, so that
\begin{equation} 
{\cal A}^a_0(\vec r,t=0)=0 \,,\quad
{\cal E}^a_m(\vec x,t=0)=0\,. 
\end{equation} 
This is illustrated in Fig.~\ref{IAfig}(b).
Since configurations of this type interpolate between a mostly dual region, with
${\cal E}^a_m(z_I)={\cal B}^a_m(z_I)$, to an anti-dual region, 
where ${\cal E}^a_m(z_{\bar I})=-{\cal B}^a_m(z_{\bar I})$, it is intuitive
that the electric field vanishes in the center.

This situation can be readily interpreted in the ${\cal A}_0=0$ gauge, 
in which the electric field is simply the time derivative of the
gauge field -- the canonical {\em momentum} in Yang-Mills field quantization.
Thus the $t=0$ magnetic state is indeed identified as 
a turning state, in which motion is momentarily stopped.
For separation $T$ comparable to the size $\rho$ the energy is finite,
with a maximum $E\sim 1/(g\rho)$.

The energy $E$ and Chern-Simons number $N_{CS}$ for either the sum or ratio 
ansatz can be calculated as a function of separation $T$ directly, 
with the hope that a parametric plot of $E(N_{CS})$ will reveal a useful
profile of the barrier as a function of this topological coordinate.

Alas, for the sum ansatz this idea produces reasonable results only for very 
large separation, $T\geq 2\rho$.
When $T$ is of the order $\rho$, the energy $E(T)$ of the turning state 
(as well as the action for the entire configuration) becomes very large, 
while the topological coordinate $N_{CS}(T)$ remains fixed. 
It is therefore obvious that this set of paths does not describe the travel 
across the ridge separating classical vacua which we want to study.
Instead, this path rises with the barrier but continues to increase as the
origin is approached, following a direction apparently orthogonal to
the topological coordinate we want to study.

\begin{figure}[h]
\hspace*{-5mm}
\begin{center}
\epsfig{file=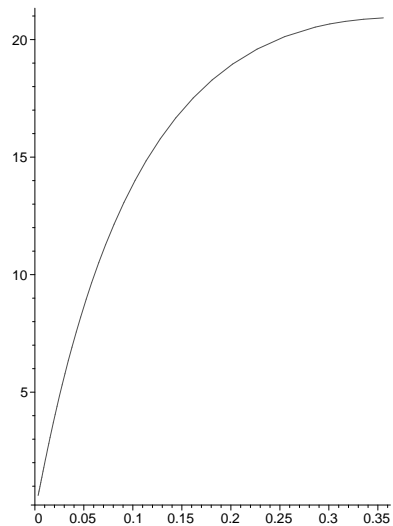, width=50mm}
\end{center}
\caption {\label{rat_path}
The normalized energy, $ER$, versus the Chern-Simons number
for the ratio ansatz.
}
\end{figure}

The ratio ansatz yields somewhat better results, with finite (and
even simple) field structure at all $T$, including the point $T=0$. 
However the results, shown in Fig.~\ref{rat_path},
indicate that this set of trial functions can only accomplish about 
one third of the journey we would like to make, in terms of the topological 
quantity $N_{CS}$.
This inadequacy will become apparent after comparison with the results 
to follow.

\subsection{The Yung Ansatz, or Going Uphill} 

As a natural set of \IA configurations, one of us \cite{Shu_qm}
suggested starting from a well-separated pair and 
going {\em downhill}, along the gradient of the action\footnote{
This can easily be done numerically, and a set of such curves
for the quantum-mechanical double well potential and
the corresponding set of \IA configurations was found in that work.}. 
Naturally, minimization of the action leads to
complete \IA annihilation and zero action.

It was shown by Yung that these configuration can generally be obtained 
from a solution of the streamline equation \cite{yung}. 
He found solutions for large separation $T\gg \rho$ and used them to derive
the next order terms in the \IA interaction, to $O(1/T^6)$. 
A clever conformal symmetry was used to reduce the
Yang-Mills problem to that of a double well potential.
The same trick was then used in the numerical solution of the streamline 
equation \cite{KhR,Verbaarschot}, in which it was observed
that the approximate ansatz suggested by Yung also happens to be  
a very accurate approximation to true solution, not only at large $T$
(as Yung intended) but in fact for {\em all} finite \IA separations $T$. 
As expected, at $T=0$ \IA annihilation occurs and
the field strength vanishes\footnote{
This is not obvious from the Yung expression; it was first found numerically.
The Yung formula's complicated result at $T=0$ is nothing but a 
pure gauge.}.

Since we take a different view of \IA configurations in this work,
we interpret a solution of the streamline equation (or Yung ansatz) as
a set of forced paths going {\em uphill} against the gradient of the
force. 
This process reaches its turning point (or state), with some maximal energy
and Chern-Simons number, and then turns back.  
Because the process proceeds uphill, unlike with other trial functions
with some arbitrary driving force, we expect that all trajectories rise 
along {\em the same path}, although those with larger $T$ go further up.

The Yung ansatz for the field configuration is rather complicated,
and is best written in matrix form: 
\begin{eqnarray} 
ig{\cal A}_{\mu}^{Yung}(x) &=& 
ig{\cal A}_{a\mu}^{Yung}(x)\frac{\tau^a}{2} \nonumber \\
&=& {\bar {\tilde y_2} \over \sqrt{\tilde y_2} }
 {R \over \sqrt{R^2} } 
 {(\bar \sigma_\mu y_1-y_1^\mu)\rho_1^2 \over y_1^2 ( y_1^2+\rho_1^2)}
 {\bar R \over\sqrt{R^2}}{\bar {\tilde y_2} \over \sqrt{\tilde y_2}}  
 \nonumber \\
&& + {(\bar \sigma_\mu y_2-y_2^\mu)\rho_2^2 \over  y_2^2+\rho_2^2}+ 
 {\rho_1\rho_2 \over z y_1^2( y_2^2+\rho_2^2)}\nonumber \\
&& \Bigg[ (\bar \sigma_\mu y_1-y_1^\mu)-
{\bar {\tilde y_2} \over \sqrt{\tilde y_2}}\nonumber \\
&& {R \over \sqrt{R^2}} (\bar \sigma_\mu y_1-y_1^\mu) 
{\bar R \over\sqrt{R^2}}{\bar {\tilde y_2} \over \sqrt{\tilde y_2}}
\Bigg] ,
\end{eqnarray} 
where $z$ is related to the conformal-invariant distance,
$(R^2+\rho_1^2+\rho_2^2)/(\rho_1\rho_2)$. 
In the case
$\rho_1=\rho_2=\rho$, which is the only one we need, this relation reads
\begin{equation} 
z^2=\frac{R^2+2\rho^2+\sqrt{(R^2+2\rho^2)^2-4\rho^2} }{2 \rho^2} \,.
\end{equation} 
All vectors without an indicative index are SU(2) matrices
obtained by their contraction with the vector $\sigma_\mu=(1,-i\vec\tau)$, 
for example $R=x_1-x_2=R_\mu \sigma_\mu$.  
An overbar similarly denotes contraction with $\bar\sigma= (1,i\vec\tau)$. 
Note that barred and unbarred matrices always alternate, in all terms; 
this is because one index of each matrix is dotted and the other not, 
in spinor notation.
Finally, the additional coordinate is
\begin{equation} 
\tilde y_2=x_2- \frac{R \rho_2}{z\rho_1-\rho_2} \,.
\end{equation} 

Note that the first term is the instanton in the {\em singular} gauge,
the second is the anti-instanton in the {\em regular} gauge, and the third 
is a ``correction'' term.
The benefit of this representation is that the same 't Hooft symbol appears 
in all three terms, and the entire construction originates from conformal 
transformation of a spherically symmetric configuration in which $\bar I,I$ 
share the same center. 
An unfortunate feature of this expression is that time-reversal symmetry
is far from obvious, and it is not clear that the electric field at the 
mid-plane vanishes. 
However, this is in fact the case and the field at $t=0$ can be interpreted
as a turning state.

\begin{figure}[tb]
\begin{center}
\epsfig{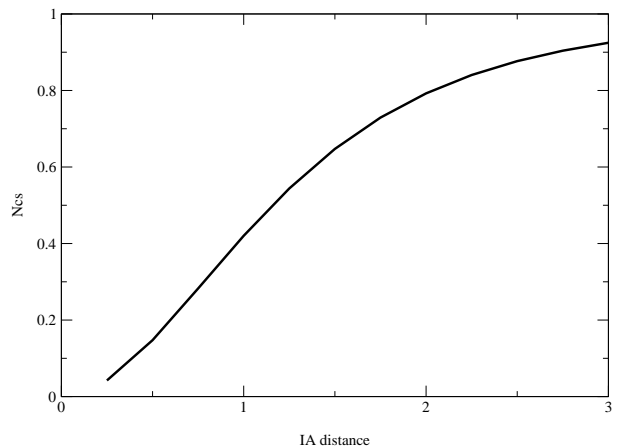}
\end{center}
\caption {\label{NCSvsR}
$N_{CS}$ versus the distance between $\bar I I$ centers $T$ in the Yung ansatz. 
}
\end{figure}

Although all three trial functions are similar at large
\IA separation $T$, they are drastically different at $T\sim \rho$.
The Yung ansatz is the only one which allows us to reasonably study 
the effects of a large change in topological number.
The variation of the Chern-Simons number of the turning state ($t=0$)
as a function of the \IA separation $T$ can be seen in
Fig.~\ref{NCSvsR}. 
In this case we scan the entire range $[0,1]$.

\begin{figure}
\begin{center}
\epsfig{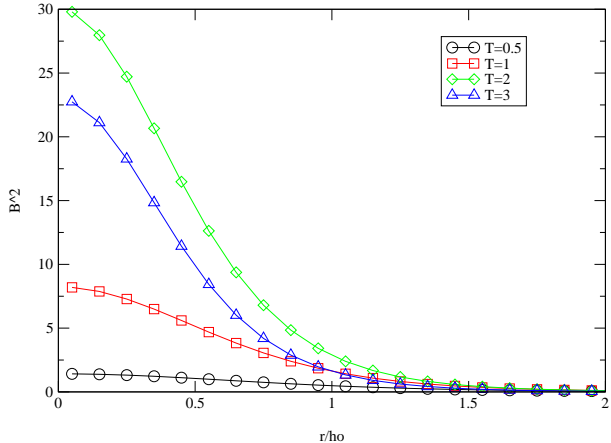}
\end{center}
\caption {\label{BB_profile}
The ${\cal B}(r)^2$ profile, not normalized, 
for the four values of the $\bar I I $
distance $T$ (in units of $\rho$) indicated in the legend.
}
\end{figure}

We now proceed with a more detailed study of the static turning states,
residing on the $t=0$ 3-plane. 
The simplest observable is the shape of the corresponding
magnetic field squared, or the energy density distribution, shown 
in Fig.~\ref{BB_profile} for few selected values of $\bar I I $
distance $T$. 
Note that the curve for $T=2$ (the most like the sphaleron)
show indeed the largest magnitude of the magnetic field.
The shape is however rather uniform. 
Note also that, unlike the case of the faulty sum and ratio trial 
functions, for smaller $T$ the field strength decreases, 
ultimately disappearing at $T=0$.  

\begin{figure}[h]
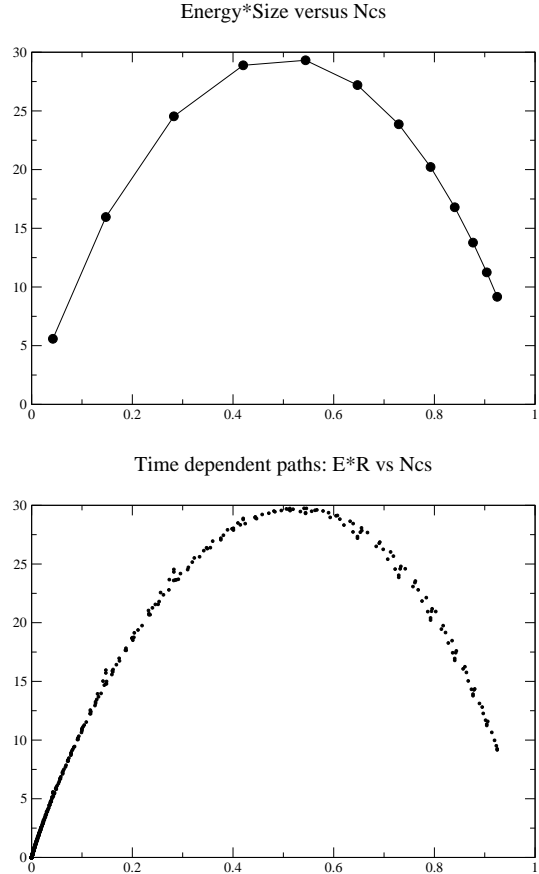

\begin{center}
\epsfig{file=fig8.eps, width=70mm}
\epsfig{file=fig9.eps, width=70mm}
\end{center}
\caption {\label{ERvsNCS}
The normalized energy, $ER$, versus the Chern-Simons number
for the Yung ansatz. Plot (a) shows the positions of the turning states
for various $T$, while (b) combines many points along the path
($t\ne0$); their small spread means that Yung ansatz is nearly
going directly uphill, thus passing via the same points for different $T$.  
}
\end{figure}

The energy and energy density of the turning state configurations is therefore
rather different for different $T$. 
However, as seen from Fig.~\ref{BB_profile}, the physical sizes of these
objects are different as well. 
As classic Yang-Mills theory has scale invariance, one may wish to make the
more natural comparison of a scale-invariant combination,
the energy times the r.m.s. radius, $R$, defined as
\begin{equation} 
R^2= {\int d^3r\, r^2 {\cal B}^2 \over \int d^3r\, {\cal B}^2}.
\end{equation} 
In these terms, the normalized energy is
\begin{equation} 
ER=\frac{1}{2}
\left[\int d^3r r^2 {\cal B}^2 \times \int d^3r {\cal B}^2\right]^{1/2}.
\end{equation} 
This quantity is plotted versus the topological charge difference in 
Fig.~\ref{ERvsNCS}, and indeed displays a parabolic-looking maximum 
near $N_{CS}=1/2$. 

Instead of only looking at the static $t=0$ (and zero electric field)
turning states, one can instead follow the (scale invariant) energy $ER$ 
and the  Chern-Simons number as a function of time $t$ along each each path. 
As expected, {\em all} the paths in Fig.~\ref{ERvsNCS}(b), for any $T$, 
actually climb nearly exactly the same cliff, as they propagate into larger
values of our topological coordinate.

\section{Turning States from Constrained Minimization}

We will now define turning states in terms of the gauge field,
which connect the Euclidean and Minkowski domains of
the field's path. 
The turning state is characterized by the
condition that the generalized momentum, which in the
${\cal A}_0=0$ gauge coincides with the chromoelectric field,
vanishes or, equivalently, that all first time derivatives of the
spatial field components are zero. 
Using the notation introduced in Section IB,
this in turn means that
$\partial_0\phi=\partial_0\alpha=\partial_0 A_1=0$ at the
time when the transition occurs. 
From now on we assume that moment to be $t=0$.

At any given time it is possible to use the special
gauge transformation, Eq.~(\ref{gauge_transform}), with a time-independent
angle $\beta$ to gauge out $A_1(r)$, still within the $A_0=0$ gauge.
At $t=0$ the energy of the field can thus be written
\begin{equation} 
\label{Estat}
E=\frac{4\pi}{g^2}\int dr \left[(\partial_r\phi)^2+\phi^2(\partial_r\alpha)^2+
\frac{(1-\phi^2)^2}{2r^2}\right] .
\end{equation} 

We now address the question of the {\em minimal}
potential energy of static Yang-Mills field, consistent with
the appropriate constraints:
(i) a fixed value of the (corrected) Chern-Simons number, and
(ii) a given value of the r.m.s. size.
The former parametrizes the position of configuration on the topological
scale, and the latter is needed to break dilatation symmetry of the problem,
which otherwise prevents any configuration of finite size from being the 
minimum of the energy.

We will break the scale invariance of the theory by setting a
requirement that the ratio
\begin{equation}
\langle r^2 \rangle =\frac{\int d^3x\, r^2 { B}^2}{\int d^3x\, { B}^2}
\end{equation}
has a particular value, $\rho^2$, for the static solution we seek.
To keep both the Chern-Simons number and mean radius constant we introduce
Lagrange multipliers and search for their minimal combination of\footnote{
Without the term introduced to fix the Chern-Simons number
or, equivalently, for zero corresponding Lagrange multiplier,
this problem would be equivalent to the SU(2) sphaleron on a 3-d sphere 
that was solved by Smilga \cite{Smilga}.}
\begin{eqnarray}
\tilde{E}&=&\frac{4\pi}{g^2}
         \! \int \! dr \left(1+\frac{r^2}{\rho^2}\right)
                \!  \left[
                          (\partial_r \phi)^2
                        + \phi^2 (\partial_r\alpha)^2
                        + \frac{(1-\phi^2)^2}{2r^2}
                  \right]\nonumber\\
&&+\frac{\eta}{2\pi}\int dr (1-\phi^2)\partial_r\alpha \,,
\end{eqnarray}
where the tilde denotes the constrained energy.
It is convenient to introduce a dimensionless variable, 
\[
\xi=2\arctan\left(\frac{r}{\rho}\right)-\frac{\pi}{2} \,,
\]
so that
\begin{eqnarray}
\tilde{E}=\frac{8\pi}{g^2}
          \int\limits_{-\pi/2}^{\pi/2} d\xi
                 &\Big[&
                          (\partial_\xi \phi)^2
                        + \phi^2 (\partial_\xi\alpha)^2
                        + \frac{(1-\phi^2)^2}{2\cos^2\xi}\nonumber\\
      &&+\kappa(1-\phi^2)\partial_\xi\alpha\Big]
\end{eqnarray}
where $\kappa=\eta\rho g^2/(32\pi^2)$.

The Euler-Lagrange equations for the remaining fields are
\begin{equation} 
\label{EL1}
\partial^2_\xi\phi-\phi(\partial_\xi\alpha)^2+
\frac{(1-\phi)^2\phi}{\cos^2\xi}+2\kappa\phi\partial_\xi\alpha=0
\end{equation} 
and
\begin{equation} 
\label{EL2}
\partial_\xi(\phi^2\partial_\xi\alpha)+\kappa\partial_\xi(1-\phi^2)=0 \,.
\end{equation} 
Finiteness of the energy requires the boundary conditions
$\phi^2(-\pi/2)=\phi^2(\pi/2)=1$.
Eq.~(\ref{EL2}) integrates to
\begin{equation} 
\partial_\xi\alpha=-\kappa\frac{1-\phi^2}{\phi^2} \,,
\end{equation} 
with a vanishing integration constant as follows from the form of
the energy. 
After the substitution of $\partial_\xi\alpha$ into Eq.~(\ref{EL1}) one has
\begin{equation} 
\partial^2_\xi\phi+\frac{(1-\phi^2)\phi}{\cos^2\xi}=
\kappa^2\frac{1-\phi^4}{\phi^3} \,.
\end{equation} 
A solution to this equation exists for $-1<\kappa<1$,
\begin{equation} 
\phi^2=1-(1-\kappa^2)\cos^2\xi  \,.
\end{equation} 
Hereafter we assume that $\phi$ is positive.

In term of the usual $r$ coordinate, we have instead
\begin{eqnarray}
\phi(r) &=& \left(1-(1-\kappa^2)\frac{4\rho^2 r^2}{(r^2+\rho^2)^2}\right)^{1/2}
\nonumber\\ 
\partial_r\alpha(r)&=&-2\kappa\frac{1-\phi^2}{\phi^2}\frac{\rho}{r^2+\rho^2}\,.
\label{solved}
\end{eqnarray}
The sphaleron solution corresponds to $\kappa=0$ and
\begin{equation}
\phi(r)=\frac{|r^2-\rho^2|}{r^2+\rho^2}\,,\quad \alpha(r)=\pi\theta(r-\rho).
\end{equation}

\begin{figure}
\begin{center}
\epsfig{file=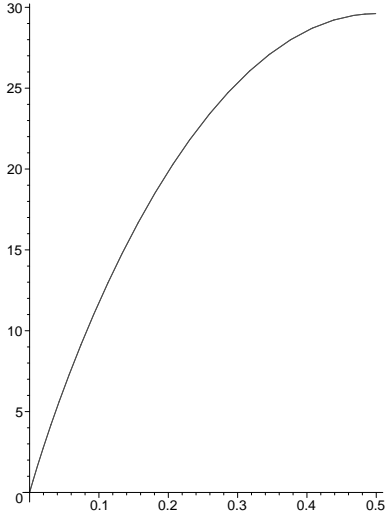, width=50mm}
\end{center}
\caption {\label{dima's_profile}
The potential energy
$E_{B}$ versus $\tilde{N}_{CS}$, for the analytic turning state
solution of Eq.~(\ref{solved}).
}
\end{figure}

For any $\kappa$ and mean squared radius
$\langle r^2 \rangle=\rho^2$, the potential energy density is
\begin{equation}
\frac{1}{2} B^2 = 24 \rho^4 \frac{(1-\kappa^2)^2}{(r^2+\rho^2)^4} \,,
\end{equation}
the integral of which is the potential (magnetic) energy of the 
static configuration,
\begin{equation}
E_B = 3\pi^2 \frac{(1-\kappa^2)^2}{g^2\rho} \,. 
\end{equation}
The corrected Chern-Simons number, computed from the first term of 
Eq.~(\ref{csdef}), is
\begin{equation}
\tilde{N}_{CS}= \frac{1}{4}{\rm sign}(\kappa)(2+|\kappa|)(1-|\kappa|)^2\,.
\end{equation}
Figure~\ref{dima's_profile} shows the profile of the potential energy
$E_{B}$ versus $\tilde{N}_{CS}$. 
It is very similar, although not identical, to the findings of 
the preceding section (see Fig.~\ref{ERvsNCS}) 
where Yung's ansatz was used for forced paths.

\section{Explosions of the Turning States: Analytic Treatment}

We are now going to use the static field configuration,
found in previous section, as an initial condition for
real-time, Minkowski evolution of the gauge field. 
Let us first consider the equations of motion in the 1+1 dimensional
dynamical system.
Variation of the action, Eq.~(\ref{action}), gives
\begin{eqnarray}
\label{ELt1}
\partial_\mu\partial^\mu\phi+\phi(\partial_\mu\alpha-A_\mu)^2
+\frac{(1-\phi^2)\phi}{r^2}&=&0
\\
\label{ELt2}
\partial^\mu\left[\phi^2\left(\partial_\mu\alpha-A_\mu\right)\right] &=& 0
\\
\phi^2(\partial_1\alpha-A_1)-
\partial_0\left[\frac{r^2}{2}(\partial_0A_1-\partial_1A_0)\right]&=&0
\nonumber\\
\label{ELt3}
\phi^2(\partial_0\alpha-A_0)-
\partial_1\left[\frac{r^2}{2}(\partial_0A_1-\partial_1A_0)\right]&=&0.
\end{eqnarray}
The solution of Eq.~(\ref{ELt2}) has the form
\begin{eqnarray}
\phi^2(\partial_0\alpha-A_0) &=& -\partial_1\psi \nonumber\\
\phi^2(\partial_1\alpha-A_1) &=& -\partial_0\psi \,,
\end{eqnarray}
where $\psi(r,t)$ is an arbitrary smooth function.
Eqs.~(\ref{ELt3}) are consistent with this solution if
\begin{equation} 
\partial_0A_1-\partial_1A_0=-\frac{2\psi}{r^2}
\end{equation} 

Now, combining Eq.~(\ref{ELt2}) and Eqs.~(\ref{ELt3}) one has
\begin{equation} 
\label{psi_rt}
\partial^\mu\left(\frac{\partial_\mu\psi}{\phi^2}\right)=
\partial_0A_1-\partial_1A_0=\frac{2\psi}{r^2},
\end{equation} 
which can be viewed as a necessary and sufficient condition
for $\psi$ to be a solution for Eq.~(\ref{ELt2}) and Eqs.~(\ref{ELt3})
simultaneously.
Eq.~(\ref{ELt1}) is now
\begin{equation} 
\label{phi_rt}
\partial_\mu\partial^\mu\phi-\frac{(\partial_\mu\psi)^2}{\phi^3}
+\frac{(1-\phi^2)\phi}{r^2}=0\,.
\end{equation} 

The initial conditions for Eqs.~(\ref{psi_rt}) and (\ref{phi_rt}) are
\begin{eqnarray}
\phi(r,0)&=&\phi(r)\,,
\nonumber\\
\partial_0\phi(r,t)|_{t=0}&=&0\,,
\nonumber\\
\partial_1\psi(r,0)&=&-\phi(r)^2\partial_0\alpha(r)=0 
\Rightarrow \psi(r,0)=0,
\nonumber\\
\partial_0\psi(r,t)|_{t=0}&=&-\phi(r)^2\partial_1\alpha(r),\nonumber
\end{eqnarray}
where the $t$-independent fields on the right sides of the equations
are the static solutions of $\phi$ and $\alpha$ from the previous section.

As with static solutions, it is more convenient to discuss the time-evolution
equations in hyperbolic coordinates. 
Let us choose $\omega$ and $\tau$ such that
\begin{equation}
\label{tau-omega}
r=\frac{\rho\cos\omega}{\cos\tau-\sin\omega} \,, 
\quad t=\frac{\rho\sin\tau}{\cos\tau-\sin\omega} \,.
\end{equation}
The physical domain of $0<r<\infty$ and $-\infty<t<\infty$ 
is covered by $-\pi/2<\omega<\pi/2$ and $-\pi/2+\omega<\tau<\pi/2-\omega$.
For $t>0$, the corresponding domain is
$-\pi/2<\omega<\pi/2$ and $0<\tau<\pi/2-\omega$. 
This change of variables (\ref{tau-omega}) is a conformal one.

In the new variables Eqs.~(\ref{psi_rt}) and (\ref{phi_rt}) become\footnote{
One can find a discussion of Eqs.~(\ref{eqs_ot}) and some of
its solutions in \cite{FKS}.}
\begin{eqnarray}
\label{eqs_ot}
-\pt^2\phi+\po^2\phi-\frac{(\pt\psi)^2-(\po\psi)^2}{\phi^3}
+\frac{(1-\phi^2)\phi}{\cos^2\omega} &=& 0 
\nonumber\\
-\pt\frac{\pt\psi}{\phi^2}+\po\frac{\po\psi}{\phi^2}-
\frac{2\psi}{\cos^2\omega} &=& 0 \,.
\end{eqnarray}

Before solving these equations let us note that it is possible to predict
the large-$t$ behavior of gauge field from the form of 
the conformal transformation (\ref{tau-omega}). 
Indeed, the $t\rightarrow\infty$ limit corresponds to the line
$\tau=\pi/2-\omega$ on the $(\omega,\tau)$ plane. 
If one now takes the limit $|r-t|\rightarrow\infty$ 
(regardless of the limit for $|r-t|/t$),
the position on $(\omega,\tau)$ plane is either 
$\omega\rightarrow -\pi/2\,,\,\tau\rightarrow 0$ or $\omega\rightarrow \pi/2
\,,\, \tau\rightarrow\pi$. 
This means that the entire line $\tau=\pi/2-\omega$
corresponds to space-time points with finite differences between
$r$ and $t$ and, therefore, if $\phi$ and $\psi$ are smooth functions
of $\omega$ and $\tau$, then for asymptotic times the field is concentrated
near the $r=t$ line.
This corresponds to the fields expanding as a thin shell in space.

We must now supply Eqs.~(\ref{eqs_ot}) with initial conditions, which are
\begin{eqnarray}
\label{init}
\phi(\omega,\tau=0)^2 &=& 1-(1-\kappa^2)\cos^2\omega
\nonumber\\
\pt\phi(\omega,\tau)|_{\tau=0} &=& 0
\nonumber\\
\psi(\omega,\tau=0) &=& 0
\nonumber\\
\pt\psi(\omega,\tau)|_{\tau=0} &=& \frac{\rho}{1-\sin\omega}
\partial_t\psi(\omega,\tau)|_{t=0} 
\nonumber\\
&=&\kappa(1-\kappa^2)\cos^2\omega \,.
\end{eqnarray}
One of the solutions of Eqs.~(\ref{eqs_ot}), first found in 1977 by
L\"uscher \cite{Luscher} and Schechter \cite{Schechter}, is
\begin{eqnarray}
\phi(\omega,\tau)^2 &=& 1-(1-q^2(\tau))\cos^2\omega \nonumber\\
\psi(\omega,\tau) &=& \frac{\dot{q}(\tau)}{2}\cos^2\omega \,,
\end{eqnarray}
with a function $q(\tau)$ that satisfies
\begin{equation}
\label{ddotq}
\ddot{q}-2q(1-q^2)=0 \,.
\end{equation}
This is the equation for a one-dimensional particle
moving in double-well potential of the form $U(q)=(1-q^2)^2/2$.

We now have to check that the L\"uscher-Schechter solution
satisfies the initial conditions, (\ref{init}).
This is indeed the case if one identifies $q(0)=\kappa$ 
and takes $\dot{q}(0)=0$.
For the initial condition of this type
({\em i.e.} for energy $\varepsilon=\dot{q}^2/2+U(q)<1/2$),
the solution of Eq.~(\ref{ddotq}) is
\begin{equation}
\label{solution}
q(\tau)=\tilde{q}\dn\left(\tilde{q}(\tau-\tau_0), k\right),
\end{equation}
where $\dn$ is Jacobi's function and
$\tilde{q}=\sqrt{2-\kappa^2}$ is the second stopping point
for a particle in the potential $U(q)$.
We have also defined
\[
k^2=2\frac{1-\kappa^2}{2-\kappa^2} \quad {\rm and}\quad
\tau_0\tilde{q}=\frac{T}{2} \,,
\]
where $T$, the period of oscillations in the potential $U(q)$, is 
$T=2K(k)$, with $K(k)$ being the complete elliptic integral of the first kind.
The idea is, of course, that ``oscillations'' in $\tau$ begin
from the rest point, close to $\tau = 0$.

Let us now look at several properties of the solution for large times.
The solution (\ref{solution}) is apparently regular in the $(\omega, \tau)$
plane, and therefore for large times the field is concentrated near $r=t$. 
At asymptotic times the energy density, $e(r,t)$, is given by
\begin{equation}
4\pi e(r,t) = \frac{8\pi}{g^2\rho^2}(1-\kappa^2)^2
\left(\frac{\rho^2}{\rho^2+(r-t)^2}\right)^3 \,.
\end{equation}
The change in topological charge is
\begin{eqnarray}
\Delta Q &=& \int\limits_0^{\infty} d^3x dt \,\partial_\mu K^\mu\nonumber\\
&=& \frac{1}{2\pi}\int dr dt
\left[-\partial^2_t\psi+\partial^2_r\psi-\frac{2\psi}{r^2}\right]\nonumber\\
&=& \frac{\pi}{2}\kappa(3-\kappa^2)
-{\rm sign}(\kappa)\arccos\left(
\frac{\mbox{\rm cn}(\tilde{q}\pi, k)}{\dn(\tilde{q}\pi, k)}\right) .
\end{eqnarray}
The evolution of $\tilde{N}_{CS}$ begins from time $t=0$, where
\begin{equation}
\tilde{N}_{CS}(0)=\frac{1}{4}{\rm sign}(\kappa)(1-|\kappa|)^2(2+|\kappa|) \,,
\end{equation}
and as $t\rightarrow\infty$ its limit is
$\tilde{N}_{CS}(\infty)=\tilde{N}_{CS}(0)+\Delta Q$.

We now estimate number of gluons produced by the described evolution. 
In $\phi,\psi$ language the chromoelectric and chromomagnetic
fields are
\begin{eqnarray}
{ E}^a_j  &=&
\frac{1}{r} \left(\partial_t\phi\sin\alpha -
\frac{\partial_r\psi\cos\alpha}{\phi}\right)\Theta^a_j \nonumber\\
&&+ \frac{1}{r} \left(\partial_t\phi\cos\alpha +
\frac{\partial_r\psi\sin\alpha}{\phi}\right)\Pi^a_j +
\frac{2\psi}{r^2}\Sigma^a_j \,,
\label{electric2}
\end{eqnarray}
\begin{eqnarray}
{ B}^a_j &=&
-\frac{1}{r} \left(\partial_r\phi\cos\alpha +
\frac{\partial_t\psi\sin\alpha}{\phi}\right)\Theta^a_j \nonumber \\
&& + \frac{1}{r}\left(\partial_r\phi\sin\alpha -
\frac{\partial_t\psi\cos\alpha}{\phi}\right)\Pi^a_j +
\frac{1-\phi^2}{r^2}\Sigma^a_j \,.
\label{magnetic2}
\end{eqnarray}
Terms proportional to $\Sigma^a_j$ are longitudinal and die out as 
$t\rightarrow\infty$. 
The remainder is a purely transverse field.
Now let us take into account that for $t\rightarrow\infty$,
$\partial_r\phi\rightarrow\partial_t\phi$, and the same for $\psi$.
Therefore
\begin{eqnarray}
{ E}^a_j &\rightarrow&
\frac{1}{r} \left(\partial_r\phi\sin\alpha -
\frac{\partial_r\psi\cos\alpha}{\phi}\right)\Theta^a_j  \nonumber \\
&&+   \frac{1}{r} \left(\partial_r\phi\cos\alpha +
\frac{\partial_r\psi\sin\alpha}{\phi}\right)\Pi^a_j \,,
\label{electric3}
\end{eqnarray}
\begin{eqnarray}
{ B}^a_j &\rightarrow&
-\frac{1}{r} \left(\partial_r\phi\cos\alpha +
\frac{\partial_r\psi\sin\alpha}{\phi}\right)\Theta^a_j  \nonumber \\
&&+   \frac{1}{r} \left(\partial_r\phi\sin\alpha -
\frac{\partial_r\psi\cos\alpha}{\phi}\right)\Pi^a_j \,.
\label{magnetic3}
\end{eqnarray}
The main result becomes apparent when we choose a gauge where
\[\phi\partial_r\phi\cos\alpha + \partial_r\psi\sin\alpha=0\,, \]
in which 
\begin{eqnarray}
{ E}^a_j &\rightarrow&
\frac{1}{r}
\sqrt{\frac{(\partial_r\psi)^2}{\phi^2}+(\partial_r\phi)^2}\Theta^a_j 
\nonumber\\ 
&\rightarrow&
\frac{1-\kappa^2}{r\rho}\left(\frac{\rho^2}{\rho^2+(r-t)^2}\right)^{3/2}
\Theta^a_j \,,
\label{electric4}
\end{eqnarray}
\begin{equation}
{ B}^a_j \rightarrow
\frac{1-\kappa^2}{r\rho}\left(\frac{\rho^2}{\rho^2+(r-t)^2}\right)^{3/2}
\Pi^a_j \,.
\label{magnetic4}
\end{equation}

We now perform a fourier transform, finding
\begin{eqnarray}
{ E}^a_j(\vec{k})
&=& 4\pi\rho(1-\kappa^2)K_1(\omega\rho)\Theta^a_j\nonumber\\
{ B}^a_j(\vec{k})
&=& 4\pi\rho(1-\kappa^2)K_1(\omega\rho)\Pi^a_j\,,
\end{eqnarray}
where $\Theta^a_j$ and $\Pi^a_j$ are the color/space 
projectors in momentum space analogous to those in coordinate space
(\ref{projectionops}), the frequency $\omega = |\vec{k}|$,
and $K_1$ is a Bessel function.
One can easily verify that ${ B}^a_j=\epsilon_{jlm}k_l{ E}^a_m/k$, as 
is required for a radiation field. 

\section{Explosions of the Turning States:  Numerical Studies}

In the previous section we studied the asymptotic behavior of 
turning states constrained by sphaleron size and Chern-Simons number.
In this section we consider step-by-step evolution of the turning
states, a numerical analysis similar to sphaleron decay in electroweak
theory \cite{HK,Zad}.
The numerical approach naturally allows for mathematical flexibility,
and it is used to consider the decay of static states which replace the
unphysical power-law behavior of the fields at large distance with
a phenomenologically more appropriate exponential tail.
The classical field configurations are thus fixed in size indirectly by a 
mass parameter, a constraint which is subsequentially
relaxed as the state quickly decays into free-streaming gluons.

As a result of scale invariance, the QCD instanton is of indeterminate
size.
While it is clear from phenomenology and lattice studies that the instanton
vacuum favors a somewhat narrow size distribution centered at $\bar\rho
\simeq 0.3$ fm, the reason for this is yet unknown, although
it is presumably due to interactions between instantons.
It is thus natural that related classical objects, born in some way from
the excitation of instantons, share a similar size.
We arrange this by introducing a phenomenological gluon mass term in 
the initial configuration which is promptly relaxed as this unstable
configuration begins to decay.
We stress that the relaxation of this size constraint does not initiate
the explosion; we will show that the turning state is an unstable 
configuration regardless of the mass term's presence.

The similarity between this procedure and electroweak sphaleron decay is 
clear, but not mathematically continuous.
For the Higgs mechanism, the sphaleron size is constrained by the scalar
vacuum condensate which introduces an effective mass for the gauge fields.
This condensate vanishes at the sphaleron center, where the classical
gauge field action is maximal.
This feature persists in the limit of infinite gauge-Higgs coupling, 
when the Higgs field is fixed at its VEV constant for all other points in space.
Inserting a mass term for the gauge fields by hand, as we do here for QCD,
is thus very similar to this infinite coupling limit of electroweak dynamics,
the only difference being that the mass is finite in {\it all} of space.
This leads to a difference in field behavior at the origin.

We begin with the Yang-Mills action, with a phenomenological mass term 
added, and will look for static solutions with
spherical symmetry in Minkowski space.
The action is written:
\begin{eqnarray}
S &=& \frac{4\pi}{g^2}\int dt dr \Bigg\{ \dot{\phi}_1^2 + \dot{\phi}_2^2 
+ \frac{1}{2} r^2 \dot{A_1}^2 - (\phi_1')^2 - (\phi_2')^2 \nonumber\\
&&-\frac{\left(1-\phi_1^2 - \phi_2^2\right)^2}{2 r^2} 
- 2 A_1\left(\phi_1\phi_2' - \phi_1\phi_2'\right) 
\label{gcaction} \\ &&
- A_1^2\left( \phi_1^2 + \phi_2^2 \right) 
- m^2 \left[ \left(1+\phi_1\right)^2 + 
\phi_2^2 + \frac{1}{2} r^2 A_1^2 \right] \Bigg\}. \nonumber
\end{eqnarray}
As before, we have taken a spherical ansatz similar to Witten's 
\cite{Witten_ans} for the gauge field.
Following the notations of Eq.~(\ref{sph_ansatz}),
\begin{equation}
A = \frac{1-\phi_1}{r} \,, \quad
B = \frac{\phi_2}{r}  \,, \quad
C = A_1 \,,  \quad
D = A_0  \,,
\end{equation}
where all scalar functions depend on both $r$ and $t$.
We initially work in the temporal gauge, where $A_0(r,t) = 0$.

The equations of motion are easily obtained, and we find
\begin{eqnarray}
&& \ddot{\phi_1} - \phi_1'' + \frac{\phi_1}{r^2}\left(\phi_1^2+\phi_2^2-1\right)
+ \phi_1 A_1^2 - 2 \phi_2' A_1 \nonumber\\
&& \qquad - \phi_2 A_1' + m^2\left(\phi_1 + 1\right) = 0 \,,
\nonumber \\
&& \ddot{\phi_2} - \phi_2'' + \frac{\phi_2}{r^2}\left(\phi_1^2+\phi_2^2-1\right)
+ \phi_2 A_1^2 + 2 \phi_1' A_1 \nonumber\\
&& \qquad + \phi_1 A_1' + m^2 \phi_2 = 0 \,,
\nonumber \\
&& \ddot{A_1} + \frac{2}{r^2} A_1 \left( \phi_1^2 + \phi_2^2 \right)
+ \frac{2}{r^2} \left( \phi_1' \phi_2 - \phi_1\phi_2' \right)
\nonumber\\ && \qquad 
+ m^2 A_1  = 0 \,.
\label{eom}
\end{eqnarray}

\begin{figure}[b]
\begin{center}
\epsfig{file=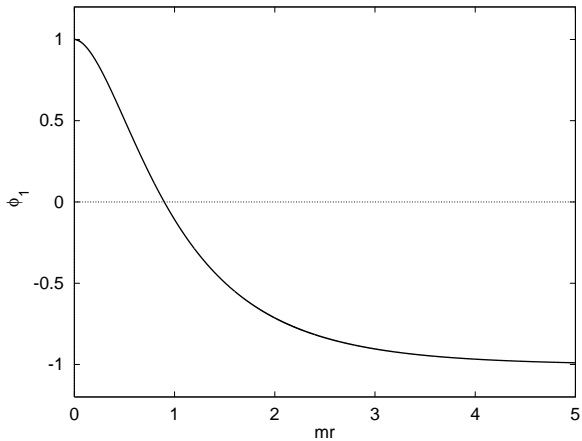, width=80mm}
\end{center}
\caption{
The static turning state solution, $\phi_1(r,0)$.
}\label{fafig}
\end{figure}

In the static limit, we seek a purely magnetic turning state solution.
One can be found for the $\phi_1$ field component from the equation
\begin{equation}
\phi_1'' - \frac{\phi_1}{x^2}\left( \phi_1^2 - 1\right) 
- \left(\phi_1+1\right) = 0 \,,
\end{equation}
which is simply Eqs.~(\ref{eom}) with $\phi_2 = A_1 = 0$ and written
in terms of the dimensionless variable $x \equiv mr$.
With the boundary conditions
\begin{eqnarray}
&\phi_1(x) \rightarrow 1 \:& {\rm as} \: x\rightarrow 0 \,, \nonumber\\
&\phi_1(x) \rightarrow -1 \:& {\rm as} \: x\rightarrow \infty \,, \nonumber
\end{eqnarray}
A numerical solution is easily obtained and shown in Fig.~\ref{fafig}.
This is quite similar to the approximate electroweak solution found by 
Klinkhammer and Manton \cite{KM} in the limit of infinite Higgs self-coupling.
The primary difference is the behavior near the origin, which in the QCD
case involves a logarithm for $x \ll 1$:
\begin{equation}
\phi_1(x) = 1 + \frac{2}{3} x^2 \ln x - \alpha x^2 + {\cal O}
\left( x^4 \ln x\right) \,,
\end{equation}
where $\alpha = 1.98$ was determined numerically.
Defining the turning state's size as the radius at which the profile is at
half its maximum ({\it i.e.} where it crosses the origin), we find
$m\rho \simeq 0.9$.  
We match this to the size of the average instanton and find
\begin{equation}
m = 0.9 \rho^{-1} \simeq 540 \,{\rm MeV}.
\end{equation}

Before we discuss the decay of this configuration we must find the 
unstable modes orthogonal to it which determine the ``downhill'' directions 
in field space.
This is done by solving eigenvalue equations for fluctuations in
the fields $\phi_2$ and $A_1$ in the presence of the turning state 
configuration.

We take the terms linear in $\phi_2$ and $A_1$ from Eqs.~(\ref{eom}) and 
require
\begin{equation}
\ddot{\phi_2}(x,t) = - \omega^2 \phi_2(x,t) \,,\quad
\ddot{A_1}(x,t) = - \omega^2 A_1(x,t)\,.
\end{equation}
We then have the eigenvalue equations:
\begin{eqnarray}
\phi_2'' + \left(\Omega^2 - 1 + \frac{1}{x^2}\right) \phi_2
+ 2 \phi_1' A_1 + \phi_1 A_1'  &=& 0 \,,\nonumber\\
\left( \Omega^2 - 1 - \frac{2}{x^2}\phi_1^2 \right) A_1
- \frac{2}{x^2} \left( \phi_1' \phi_2 - \phi_1\phi_2' \right) &=& 0 \,,
\label{unst_eom}
\end{eqnarray}
where $\phi_1$ is the classical solution in Fig.~\ref{fafig} and the 
dimensionless frequency is $\Omega = \omega/m$.
The longitudinal field component may be eliminated with
\begin{equation}
A_1 = \frac{2\left( \phi_1' \phi_2 - \phi_1 \phi_2'\right)}
{\left(\Omega^2-1\right)x^2 - 2 \phi_1^2}\,.
\end{equation}
Substituting this into the first of Eqs.~(\ref{unst_eom}), we find
the behavior near the origin:
\begin{equation}
\phi_2(x) = c x^{ \frac{1}{2}\left(1-\sqrt{1-8y}\right)} \,,
\end{equation}
where 
\[
y = \frac{1+\Omega^2}{1-\Omega^2}
\]
and $c$ is an arbitrary normalization.
Both fields vanish as $x\rightarrow\infty$.

\begin{figure}[bt]
\begin{center}
\epsfig{file=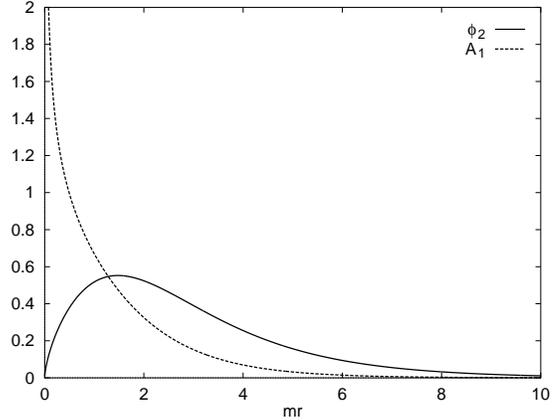, width=80mm}
\end{center}
\caption{
The unstable eigenmodes, $\phi_2(r)$ and $A_1(r)$, arbitrarily normalized.
}\label{unstfig}
\end{figure}

We have solved these equations numerically, finding
the wave functions plotted in Fig.~\ref{unstfig} with the frequency
\begin{equation}
\omega^2 = - 3.4 m^2 \,.
\end{equation}
The function $A_1$ is logarithmically divergent at the origin, reflecting
the difference between this massive model and the electroweak case
\cite{AKY2} in 
which the Higgs always vanishes at the origin\footnote{The dominant unstable
mode in the electroweak as $\lambda \rightarrow \infty$ is the 
scalar field component orthogonal to the condensate \cite{AKY2}.  
This mode is absent here.}.
There is therefore no smooth continuation between this model and the 
electroweak in the limit of large coupling.

These solutions for the unstable modes, which along with the classical
$\phi_1(r)$ complete our initial conditions, were put on a lattice with spacing
$\Delta x = 0.01$ and evolved at time steps of $\Delta \tau = 5\times 10^{-4}$,
where $\tau \equiv mt$.
The unstable modes, acting as a small push to properly initiate the decay,
were normalized as
\begin{equation}
\int dx \left( \phi_2(x)^2 + 2 x^2 A_1(x)^2 \right) = 5\times 10^{-3} \,.
\end{equation}
Coincident with this push we set $m=0$, in effect turning off the mass 
term, since here we are interested in the turning state decaying into 
the vacuum where no such term is motivated.
Although this effectively removes the size constraint on
$\phi_1(r,0)$, the subsequent dynamical expansion is a result of the 
push in the unstable directions of this saddle-point solution rather than
an inflation of the classical field configuration.

Once the real-time solutions to Eqs.~(\ref{eom}) with $m=0$
have been found at a given time step, the total energy is readily computed as
\begin{eqnarray}
E = \frac{4\pi}{g^2}\int dr \Bigg[&& \dot{\phi}_1^2 + \dot{\phi}_2^2 
+ \frac{1}{2} r^2 \dot{A_1}^2 + (\phi_1')^2 + (\phi_2')^2 \nonumber\\
&&+\frac{\left(1-\phi_1^2 - \phi_2^2\right)^2}{2 r^2} 
- 2 A_1\left(\phi_1\phi_2' - \phi_1\phi_2'\right) 
\nonumber\\ &&
+ A_1^2\left( \phi_1^2 + \phi_2^2 \right)  \Bigg]
\label{gceden}
\end{eqnarray}
At every time step it was verified that the energy remains equal to that
of the initial state.
Taking the instanton vacuum value of
\begin{equation}
\frac{8 \pi^2}{g^2} = 12 \,,
\end{equation}
the total energy of the decaying object was calculated and found to be 
\begin{equation}
E = 4.62 \rho^{-1} \simeq 2.8 \,{\rm GeV}.
\end{equation}

The Chern-Simons number was also computed at each time step.
In our present gauge, this is written as
\begin{equation}
N_{CS} = \frac{1}{2\pi}\! \int\!\! dr \left[ \left(1-\phi_1\right)\phi_2'
+ \phi_1' \phi_2 - \left(1-\phi_1^2-\phi_2^2\right)A_1 \right].
\label{gccsnum}
\end{equation}
The gauge invariance of changes in this quantity were verified
numerically.

\begin{figure}[bt]
\begin{center}
\epsfig{file=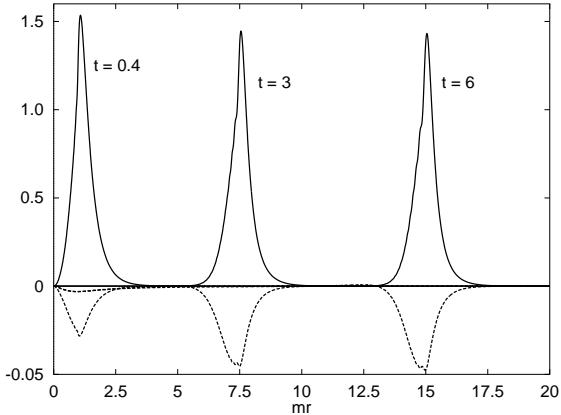, width=80mm}
\end{center}
\caption{
Energy (solid lines) and Chern-Simons number densities (dashed lines) 
for three times during the explosion, $t =$0.4, 3, and 6 fm.
(Note the scale difference between the two quantities.)
}\label{ecsfig}
\end{figure}

The energy and Chern-Simons densities, defined as the integrands of
Eqs.~(\ref{gceden}) and (\ref{gccsnum}), are shown in Fig.~\ref{ecsfig}.
The shell-like expansion is illustrated in these plots, as well as the
similarity of the energy profile at all times.
Note that these plots are in a 1+1 dimensional description, and the
corresponding three-dimensional radial density differs by a factor of
$1/r^2$.

\begin{figure}[bt]
\begin{center}
\epsfig{file=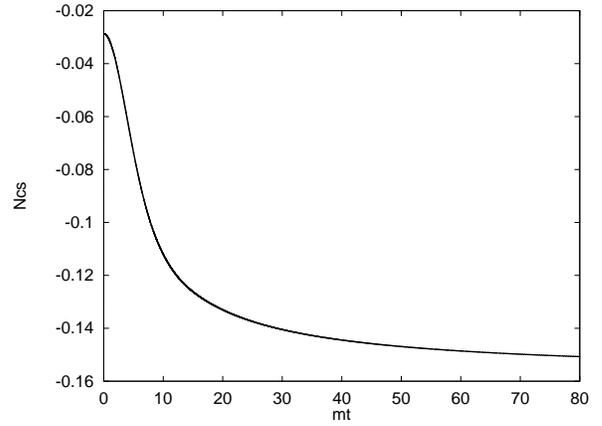, width=80mm}
\end{center}
\caption{
Change in the Chern-Simons number.
}\label{csfig}
\end{figure}

It is clear from the curves in Fig.~\ref{ecsfig} that
the Chern-Simons number changes during the decay.
As shown in Fig.~\ref{csfig},
our fields stabilize at long times with $\Delta N_{CS} \simeq 0.12$.
From this we see that the turning state does not complete an
instanton transition, which would require a return to
a state with integral Chern-Simons number.
Due to this freezing in the topology, nontrivial fermionic
solutions will accompany the resulting Yang-Mills fields.
These will be discussed elsewhere.

\begin{figure}[bt]
\begin{center}
\epsfig{file=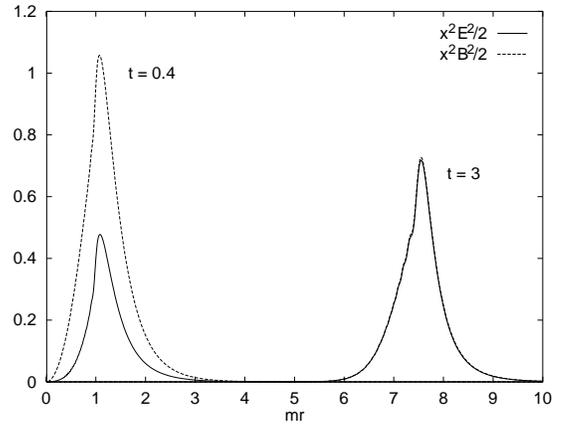, width=80mm}
\end{center}
\caption{
The electric (solid lines) and magnetic (dashed lines) fields at times
$t = 0.4$ fm and $t=3$ fm.  Plotted are
$\frac{1}{2} x^2 E(x)^2$ and 
$\frac{1}{2} x^2 B(x)^2$; their sum is the energy density.
}\label{ebfig}
\end{figure}

The transition from a purely magnetic configuration to one with equal 
electric and magnetic components is shown in Fig.~\ref{ebfig}, 
for an early and late time in the evolution.
The decay progresses rather rapidly; at $t = 1.4$ fm,
the ratio $E(r)^2/B(r)^2 \simeq 0.95$ for all $r$.
Thereafter the ratio continues to quickly approach unity.

In order to analyze the final state at late times, we
we work in a gauge in which $B(r,t) = 0$.
This requires the transformation to a new set of fields:
\begin{eqnarray}
\tilde{\phi}_1 &=& \phi_1 \cos\theta + \phi_2 \sin\theta \nonumber\\
\tilde{\phi}_2 &=& -\phi_1 \sin\theta + \phi_2 \cos\theta \nonumber\\
\tilde{A}_1 &=& A_1 - \theta' \nonumber\\
\tilde{A}_0 &=& - \dot{\theta} \,,
\end{eqnarray}
where 
\begin{equation}
\theta = \arctan\left(\frac{\phi_2}{\phi_1}\right) \,.
\end{equation}
Promptly dropping the tildes, we write the total energy in terms of 
the new fields,
\begin{eqnarray}
E &=& \frac{4\pi}{g^2} \int dr \Bigg[ \dot{\phi_1}^2 + \left(\phi_1'\right)^2
+ \frac{\left(1-\phi_1^2\right)^2}{2 r^2} \nonumber\\
&& + \frac{r^2}{2}\left(\dot{A_1}-A_0'\right)^2 
+ \phi_1^2 \left( A_1^2 + A_0^2 \right)\Bigg] \,.
\label{gaugeenergy}
\end{eqnarray}

At late times, the field strength is confined to a thin shell at radius
$r = t$.
Free, expanding field behavior is also observed numerically, such that
\begin{equation}
\dot{\Phi}(r,t) = - \Phi'(r,t)
\end{equation}
for each field $\phi_1$, $A_1$, and $A_0$.
This simplifies the equations of motion for the latter two,
\begin{eqnarray}
\ddot{A_1} - \dot{A_0}' &=& - \frac{2}{r^2} \phi_1^2 A_1 \nonumber\\
A_0'' - \dot{A_1}' &=& - \frac{2}{r}\left(A_0'-\dot{A_1}\right)
+ \frac{2}{r^2} \phi_1^2 A_1 \,,
\end{eqnarray}
in that the right hand sides of both vanish as the shell expands at large
times.
We can thus conclude that 
\begin{equation}
\dot{A_1} - A_0' = 0
\label{a10}
\end{equation}
for $mt \gg 1$.

The condition for this gauge is
\begin{equation}
\dot{A_0} \phi_1 + 2 A_0\dot{\phi_1} - A_1'\phi_1 - 2 A_1 \phi_1' = 0 \,.
\label{gaugecond}
\end{equation}
Combining this with Eq.~(\ref{a10}), we have
\begin{equation}
\phi_1\left( \dot{A_0}-A_1' \right) 
- 2\left( \phi_1' A_1 - \dot{\phi_1} A_0\right) = 0 \,,
\end{equation}
and can deduce that 
\begin{equation}
A_0 + A_1 = 0 
\label{zerooneeqn}
\end{equation}
at late times.

The contribution from $\phi_1$ will be of the form
\begin{equation}
\phi_1(r,t) = 1 + \varphi(r,t) \,,
\end{equation}
where $\varphi$ is an excitation above the vacuum.
Using the result just obtained (\ref{zerooneeqn}), 
its linearized equation of motion simplifies to 
\begin{equation}
\ddot{\varphi} - \varphi'' + \frac{2}{r^2} \varphi = 0 \,.
\end{equation}
The solution is of the form
\begin{equation}
\varphi(r,t) = \int dk \, \varphi(k) r \left[j_1(kr)+iy_1(kr)\right] \cos(kt)
\,,
\end{equation}
with fourier amplitudes $\varphi(k)$ and the spherical 
Bessel functions $j_1(z)$ and $y_1(z)$.

Although the majority of gluon radiation is carried in the $\phi_1$ field,
physical quanta also lie in small excitations of the field 
\[
\psi(x,\tau) = \phi_1(x,\tau) A_0(x,\tau) \,,
\]
which encodes oscillations between the transverse degrees of freedom.
From Eqs.~(\ref{gaugecond}) and (\ref{zerooneeqn}), we have a
wave equation,
\begin{equation}
\ddot\psi - \psi'' = 0 \,.
\end{equation}
These harmonics contribute to the energy via the final term in 
Eq.~(\ref{gaugeenergy}).

The total energy can then be written in momentum space as
\begin{equation}
E = \frac{16m}{g^2} \int dp \left[ p^2 \varphi(p)^2 
+ \psi(p)^2 \right] \,,
\label{penergy}
\end{equation}
in terms of the dimensionless momentum $p = k/m$.
The fourier amplitudes are computed from the solutions of the 
spatial fields:
\begin{eqnarray}
\varphi(p)^2 &=& \left| \int dx\, px\left[j_1(px)+i y_1(px)\right]
\varphi(x) \right|^2
\nonumber\\
\psi(p)^2 &=& \left| \int dx\, e^{ipx} \psi(x) \right|^2 \,.
\end{eqnarray}
Numerically, this expression for the energy is within 1\% of that of the 
initial configuration for all times $t \ge 3$~fm, further demonstrating
the rapid onset of free-field behavior.

\section{Production of Turning States in Heavy Ion Collisions}

\subsection{The pQCD Cutoff in Vacuum Versus Excited Matter}

Although this paper does not generally deal with phenomenological
applications, we will eventually come to an estimate of the number of
gluons produced in the explosion of a turning state. 
The resulting divergence in this number cannot be resolved without 
some explanation of the limited applicability of the classical
Yang-Mills description. 
This leads to the issue of the pQCD cutoff.

It is well known that in the QCD vacuum
the ``semi-hard'' or ``substructure scale'' 
$Q^2 \sim \, 1-2 \, {\rm GeV}^2$ is simultaneously
the {\em lower} boundary of pQCD as well as the {\em upper} 
boundary of low energy effective approaches such as chiral Lagrangians.
Furthermore, in any discussion of instanton-induced reactions it is implicitly
assumed that instantons were the primary source of that scale, and since 
they are included explicitly no other nonperturbative cutoffs are needed.
This, of course, is not strictly true as confining forces require the
final state be comprised of hadrons, but we make the usual separation of
scales and assume that final state interactions merely redistribute
the wave functions without changing the total probabilities. 

We assert, however, that the pQCD cutoff is quite different in heavy 
ion collisions.
It has been argued over the years that excited matter
might be in a Quark-Gluon Plasma (QGP) 
phase of QCD (see e.g. Ref.~\cite{Shu_80}).
Whether equilibrated or not, it is nevertheless qualitatively very 
different from the QCD vacuum: instantons are suppressed, and
there is neither confinement nor chiral symmetry breaking to set
a nonperturbative cutoff.
 
Therefore the limits on Yang-Mills field description are entirely different, 
and actually determined by much simpler phenomena.
The QGP, a plasma-like phase, screens itself perturbatively \cite{Shu_80}. 
A quasi-particle description becomes appropriate, 
in which the quarks and gluons have finite effective masses. 
In equilibrium and at high temperature these are ``thermal masses''; 
the gluon has the well-known effective mass \cite{Shu_80}
\begin{equation} 
M^2_g=\frac{g^2T^2}{2} \left(\frac{N_c}{3}+\frac{N_f}{6}\right)
\end{equation} 
where $N_c$ and $N_f$ are the number of colors and flavors, respectively.
Although this mass grows with temperature at high $T$, just above $T_c$
it is actually {\em  smaller} than the pQCD cutoff in vacuum.
Such non-monotonic behavior is confirmed by lattice thermodynamics data, 
which can be well fitted with quasiparticle masses. 

Moreover, in the ``RHIC window'', $T_c<T<3T_c$, one finds the approximately 
constant gluon and quark effective masses
\cite{LH}
\begin{equation} 
\label{Meff}
M_g\approx .4\,{\rm GeV}\,,\quad M_q\approx .3 \,{\rm GeV},
\end{equation} 
the first of which provides the cutoff of our classical treatment;
at this scale the
classical Yang-Mills action for gluons is to be modified by inclusion
of an appropriate effective Lagrangian describing such screening effects,
such as those suggested by Taylor and Wong \cite{TW}.

\subsection{Multiplicity and Spectra of Prompt Gluons}

With solutions for the fields at all times,
in both the analytic and numerical treatments of the previous two sections, 
we have analyzed the final states to determine the particle number.
While we find a similar number of produced particles from both approaches,
about four gluons produced per turning state, the energy distribution of
this prompt glue is very sensitive to the classical configuration used
to initiate the explosion.
While the field solutions found using constrained quantization (Section III)
and an effective mass (Section V) are qualitatively similar, both leading
to an expanding shell of radiation, we find a substantial difference in final 
state momentum distributions.
This can be traced to the details of the two solutions as 
$r\rightarrow\infty$, where we contrast a power-law behavior in
Eq.~(\ref{solved}) with an exponential fall-off in the solution of
Fig.~\ref{fafig}, as $\phi_1(r,0) \sim {\rm exp}(-mr)-1$.
As mentioned above, we consider the second case to be of greater physical
relevance, as the gluonic field is not massless phenomenologically 
(and even vacuum instantons ought to have exponential tails).
We now consider both results.

\begin{figure}[h]
\hspace*{-5mm}
\begin{center}
\epsfig{file=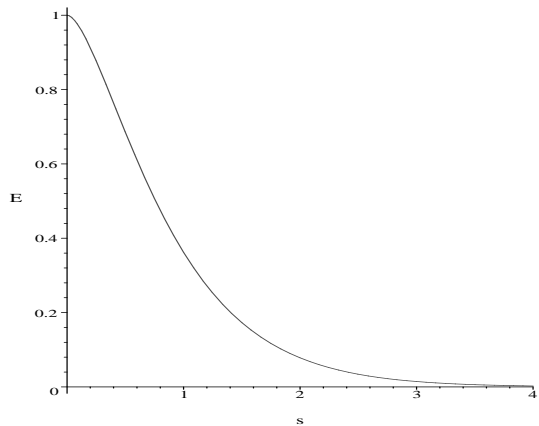, width=70mm}
\end{center}
\caption{
The energy spectrum $E(\omega)$ obtained from the analytic solution, 
in units of $\frac{16}{g^2}(1-\kappa^2)^2$, versus $s=\omega\rho$.
}\label{fig_spectrum}
\end{figure}

To find the number of gluons in each mode one compares the field 
strength in the momentum representation to those which have energy $\omega$.
In the evolution described analytically in Section IV, the
occupation number is
\begin{equation}
\nu(\vec{k})
=\frac{64\pi^2}{g^2}(1-\kappa^2)^2\frac{\rho^2}{\omega}K_1(\omega\rho)
\label{occno}
\end{equation}
and the gluon energy distribution function is
\begin{equation}
n(\omega)=\frac{32}{g^2}({1-\kappa^2})^2\omega\rho^2K_1^2(\omega\rho)
\end{equation}
The corresponding energy spectrum $E(\omega)=\omega n(\omega)$
is shown in Fig.~\ref{fig_spectrum}.

Because the fourier transforms of the fields are finite, the occupation
number (\ref{occno}) behaves as $1/\omega$ for small $\omega$.
The number of particles is thus logarithmically divergent and should be
cutoff at some low scale where the pQCD description of gluons is no longer 
valid, leading to
$
N_g \sim  log( \frac{1}{ M_g \rho}) \,.
$
As explained in the previous Subsection, in heavy ion collisions
this cutoff, identified with the gluon effective mass
of Eq.~(\ref{Meff}), is still relatively small as compared to the
typical momenta of gluons produced. 

\begin{figure}[bt]
\begin{center}
\epsfig{file=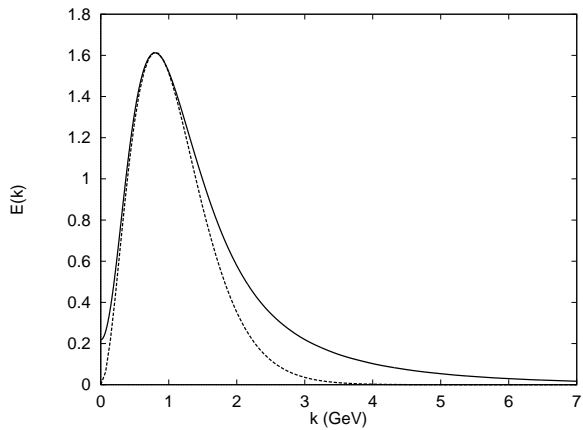, width=80mm}
\end{center}
\caption{
Energy spectrum of prompt gluons (solid line), obtained from the numerical 
solution, and a thermal distribution with $T=285$ MeV (dashed line).
}\label{edenfig}
\end{figure}

The prompt particle energy distribution was also obtained in the numerical
treatment of Section V, 
defined as the integrand of the expression in Eq.~(\ref{penergy}),
is shown in Fig.~\ref{edenfig} at a very late time in the evolution
($\tau = 50$ or $t\simeq 20$ fm).
Like the previous distribution, it is finite at the origin, but in
contrast it peaks at nonzero momentum.
For illustrative purposes it is compared with a thermal distribution 
of bosons at a temperature $T = 285$ MeV.
We note that in an equilibrated environment,
the effective mass and screening effects will only modify this profile below
$k = 0.4$ GeV, where a relatively small part of the spectrum resides.

The produced gluons are free streaming, with no mechanism for equilibration,
and yet our distribution is very similar to the thermal one for 
momenta below about 1.5 GeV.
Such a nearly thermal distribution has also been obtained from similar 
calculations using an entirely different classical field approach 
in Ref.~\cite{KV}.
One can speculate that such an ostensible equilibration may contribute to the 
the success of hydrodynamics in calculating particle spectra and elliptic 
flow at RHIC \cite{TLS}.
Our finding of four physical gluons from every decayed turning state is 
also in line with RHIC entropy production, assuming that the density of
these classical objects corresponds to that of instantons in the vacuum.
The total energy of the turning state, found above to be 2.8 GeV, 
is carried by gluons with a distribution peaked around 800 MeV.
This average can be viewed as an upper bound, since in
a more complete treatment a portion of this energy will be used for
the production of light quark pairs.
This next step will be addressed in a later publication.

\section{Conclusions}

In this work we have studied forced tunneling in pure SU(2) Yang-Mills theory.
This process, generated through the excitation of instantons in the
QCD vacuum, leads to unstable classical {\em turning states} which 
explosively decay into gluonic radiation.
These states and their decays are similar to the physics of sphalerons in
the electroweak theory.

If this semi-classical treatment of pure Yang-Mills theory is indeed 
relevant to the physics of QCD, the turning states should play a prominent
role in the production of glue in high-energy hadronic collisions.
In the case of $NN$ collisions the produced gluons will propagate into
the QCD vacuum, to be quickly recombined into secondary hadrons.
For heavy ion collisions, however, the large quantity of prompt glue 
produced from the many turning states would be released into a highly 
excited, perhaps deconfined medium.
Observable consequences in both situations were discussed recently in
Ref.~\cite{Shu_01}, and the results of this work support many of the
estimates therein.
Finally, we have obtained energy spectra for the prompt gluons that serve
as the initial state in the dynamics of a heavy ion collision and found
that, unlike the overall explosive dynamics, it is sensitive to the
details of the initial turning state profile.
We plan to further investigate phenomenological implications of the
turning states in later works, with the role of fermion production a
top priority.

{\bf Acknowledgements:}
We thank I. Zahed and R. Venugopalan for useful discussions.
This work is partially supported by the US-DOE grants Nos. DE-FG02-88ER40388
and DE-FG03-97ER41014.

\end{document}